\documentclass[longauth]{aa} 
\usepackage{graphicx}
\usepackage{txfonts}
\usepackage{amsmath,amsfonts,amssymb} 
\usepackage{natbib}
\usepackage{multicol,caption}
\usepackage{lipsum}
\usepackage{booktabs}
\usepackage{threeparttable}
\usepackage{multirow}
\usepackage{here}
\usepackage{lscape}

\bibpunct{(}{)}{;}{a}{}{,}
\usepackage[colorlinks=true,urlcolor=blue,linkcolor=blue,citecolor=blue]{hyperref}

\usepackage{threeparttable}

\usepackage{booktabs}

\usepackage{longtable}

\usepackage{ragged2e}

\usepackage{tikz,xcolor,hyperref}

\definecolor{lime}{HTML}{A6CE39}
\DeclareRobustCommand{\orcidicon}{
	\hspace{-1.5mm}
	\begin{tikzpicture}
	\draw[lime, fill=lime] (0,0) 
	circle [radius=0.16] 
	node[white] {{\fontfamily{qag}\selectfont \tiny ID}};
	\draw[white, fill=white] (-0.0625,0.095) 
	circle [radius=0.007];
	\end{tikzpicture}
	\hspace{-2.5mm}
}
\foreach \x in {A, ..., Z}{
	\expandafter\xdef\csname orcid\x\endcsname{\noexpand\href{https://orcid.org/\csname orcidauthor\x\endcsname}{\noexpand\orcidicon}}
}

\foreach \x in {a, ..., z}{
	\expandafter\xdef\csname orcid\x\endcsname{\noexpand\href{https://orcid.org/\csname orcidauthor\x\endcsname}{\noexpand\orcidicon}}
}

\begin{document}

\title{TOI-1743\,b, TOI-5799\,b, TOI-5799\,c and TOI-6223\,b: TESS discovery and validation of four super-Earth to Neptune-sized planets around M dwarfs}

\newcommand{\orcidauthorA}{0000-0002-5224-247X} 
\newcommand{\orcidauthorB}{0000-0003-1464-9276}
\newcommand{\orcidauthorC}{0000-0003-1572-7707}
\newcommand{\orcidauthorD}{0000-0002-4746-0181}
\newcommand{\orcidauthorE}{0000-0002-6523-9536}
\newcommand{\orcidauthorF}{0000-0002-1420-1837}
\newcommand{\orcidauthorG}{0000-0002-1480-9041}
\newcommand{\orcidauthorH}{0000-0002-3627-1676}
\newcommand{\orcidauthorI}{0000-0001-9911-7388}
\newcommand{\orcidauthorJ}{0000-0002-1533-9029}
\newcommand{\orcidauthorK}{0000-0001-6637-5401}
\newcommand{\orcidauthorL}{0000-0002-4903-7950}
\newcommand{\orcidauthorM}{0000-0003-1462-7739}
\newcommand{\orcidauthorN}{0000-0002-6191-459X}
\newcommand{\orcidauthorO}{0000-0003-0987-1593}
\newcommand{\orcidauthorP}{0000-0001-9087-1245}
\newcommand{\orcidauthorQ}{0000-0001-6285-9847}
\newcommand{\orcidauthorR}{}
\newcommand{\orcidauthorS}{0000-0003-0647-6133}
\newcommand{\orcidauthorT}{}
\newcommand{\orcidauthorU}{0000-0002-5741-3047}
\newcommand{\orcidauthorV}{0000-0002-2361-5812}
\newcommand{\orcidauthorW}{0000-0003-2527-1598}
\newcommand{\orcidauthorY}{0000-0002-0885-7215}
\newcommand{\orcidauthorX}{}
\newcommand{\orcidauthorZ}{0000-0002-9350-830X}
\newcommand{\orcidauthora}{0000-0001-7047-8681}
\newcommand{\orcidauthorb}{0000-0002-2072-6541}
\newcommand{\orcidauthorc}{0000-0002-0388-8004}
\newcommand{\orcidauthord}{}
\newcommand{\orcidauthore}{}
\newcommand{\orcidauthorf}{0000-0002-4909-5763}
\newcommand{\orcidauthorg}{}
\newcommand{\orcidauthorh}{0000-0001-8019-6661}
\newcommand{\orcidauthori}{}
\newcommand{\orcidauthorj}{}
\newcommand{\orcidauthorl}{0000-0001-9204-8498}
\newcommand{\orcidauthorm}{0000-0003-0030-332X}
\newcommand{\orcidauthorn}{0000-0002-0100-9184}
\newcommand{\orcidauthoro}{0000-0001-8602-6639}
\newcommand{\orcidauthorp}{0000-0001-9927-7269}
\newcommand{\orcidauthorq}{}
\newcommand{\orcidauthorr}{}
\newcommand{\orcidauthors}{}
\newcommand{\orcidauthort}{}
\newcommand{\orcidauthoru}{}
\newcommand{\orcidauthorv}{}
\newcommand{\orcidauthorw}{}
\newcommand{\orcidauthorx}{}
\newcommand{\orcidauthory}{}
\newcommand{\orcidauthorz}{}
\newcommand{\orcidauthoraa}{}
\newcommand{\orcidauthorbb}{}
\newcommand{\orcidauthorcc}{}
\newcommand{\orcidauthordd}{}
\author{
S.~Yal\c{c}{\i}nkaya\orcidA{}\inst{\ref{astro_liege},\ref{Dep_turkey},\ref{Obs_turkey},\ref{institute_ankara}} \thanks{E-mail: \color{blue}selcuk\_yalcinkaya@yahoo.com}
\and K.~Barkaoui\orcidB{}\inst{\ref{IAC_Laguna},\ref{astro_liege},\ref{MIT}} \thanks{E-mail: \color{blue}khalid.barkaoui@uliege.be}
\and \"O.~Ba\c{s}t\"urk\orcidD{}\inst{\ref{Dep_turkey},\ref{Obs_turkey}}  
\and M.~Gillon\orcidM{}\inst{\ref{astro_liege}}  
\and F.J.~Pozuelos\orcidC{}\inst{\ref{iaa}} 
\and M.~Timmermans\inst{\ref{astro_liege},\ref{birmingham}}  %
\and B.V.~Rackham\orcidH{}\inst{\ref{MIT},\ref{Kavli_MIT}} 
\and A.J.~Burgasser\orcidE{}\inst{\ref{UCSD}}  
\and P.~Mistry\orcidL{}\inst{\ref{New_south_wales}}  
\and A.~Peláez-Torres\orcidl{}\inst{\ref{iaa}} 
\and G.~Morello\orcidS{}\inst{\ref{iaa},\ref{INAF_Pa}} 
\and E.K.~Pass\orcidJ{}\inst{\ref{Kavli_MIT},\ref{Harvard_USA}}  
\and A.Bieryla\orcidK{}\inst{\ref{Harvard_USA}} 
\and D.W.~Latham\orcidI{}\inst{\ref{Harvard_USA}} 
\and K.A.~Collins\inst{\ref{Harvard_USA}}  
\and F.~Akar\inst{\ref{institute_ankara}}  %
\and Z.~Benkhaldoun\orcidQ{}\inst{\ref{ouca}} 
\and A.~Burdanov\inst{\ref{MIT}} 
\and J.~Brande\orcidb{}\inst{\ref{Univ_Kansas}} 
\and D.~R.~Ciardi\orcidU{}\inst{\ref{Caltech_IPAC}} 
\and C.A.~Clark\orcidV{}\inst{\ref{Caltech_IPAC}} 
\and E.~Ducrot\inst{\ref{Paris_Region},\ref{cea}} 
\and J.~de~Wit\inst{\ref{MIT}} 
\and B.O.~Demory\inst{\ref{unibe}} 
\and E.M.~Esmer\orcidN{}\inst{\ref{Dep_turkey},\ref{st_louis}} 
\and M.E.~Everett\orcidY{}\inst{\ref{NSF_NOIRLab}} 
\and G.~Fern\'andez-Rodriguez\inst{\ref{IAC_Laguna},\ref{Univ_LaLaguna}} 
\and A.~Fukui\orcidf{}\inst{\ref{Univ_tokyo},\ref{IAC_Laguna}}  
\and M.~Ghachoui\inst{\ref{ouca}} 
\and E.A.~Gilbert\orcidc{}\inst{\ref{Jet_Prop_Lab}} 
\and E.~Girardin\inst{\ref{Grand_Obs}}  
\and Y.~G\'omez~Maqueo~Chew\inst{\ref{unamcu}} 
\and K.~Ikuta\inst{\ref{Dep_Multi_Dsci_Japan}} 
\and K.~Isogai\inst{\ref{Okayama_Japan},\ref{Dep_Multi_Dsci_Japan}} 
\and M.J.~Hooton\orcidm{}\inst{\ref{Cavendish}} 
\and M.~Jafariyazani\orcidh{}\inst{\ref{Ames_NASA}} 
\and E.~Jehin\inst{\ref{star_liege}} 
\and J.M.~Jenkins\inst{\ref{Ames_NASA}} 
\and P.R.~Karpoor\orcidG{}\inst{\ref{UCSD}}  
\and Y.~Kawai\inst{\ref{Dep_Multi_Dsci_Japan}}   
\and K.~Kawauchi\inst{\ref{Ritsumeikan_Japan}}  
\and A.~Khandelwal\inst{\ref{unamcu}} 
\and A.C.~Kutluay\inst{\ref{institute_ankara}}  %
\and G.~Lacedelli\inst{\ref{IAC_Laguna}}  
\and M.~Lendl\inst{\ref{geneva}} 
\and M.B.~Lund\orcidW{}\inst{\ref{Caltech_IPAC}} 
\and F.~Murgas\orcidP{}\inst{\ref{IAC_Laguna},\ref{Univ_LaLaguna}} 
\and N.~Narita\inst{\ref{Univ_tokyo},\ref{Astro_tokyo},\ref{IAC_Laguna}} 
\and E.~Palle\orcidO{}\inst{\ref{IAC_Laguna},\ref{Univ_LaLaguna}} 
\and P.P.~Pedersen\inst{\ref{Cavendish},\ref{ETH_Zur_Queloz}} 
\and I.~Plauchu-Frayn\inst{\ref{uname}} 
\and A.S.~Polanski\orcida{}\inst{\ref{Lowell_Obs},\ref{Univ_Kansas}} 
\and D.~Queloz\inst{\ref{Cavendish},\ref{ETH_Zur_Queloz}} 
\and U.~Schroffenegger\inst{\ref{unibe}} 
\and R.P.~Schwarz\inst{\ref{Harvard_USA}}  
\and A.~Shporer\inst{\ref{Kavli_MIT}}  
\and E.~Softich\orcidF{}\inst{\ref{UCSD}}  
\and S.~Sohy\inst{\ref{star_liege}} 
\and A.~Soubkiou\inst{\ref{astro_liege},\ref{star_liege}}    
\and G.~Srdoc\inst{\ref{Kotiza_Obs}} 
\and I.A.~Strakhov\orcidS{}\inst{\ref{Sternberg_Ast}} 
\and A.H.M.J.~Triaud\inst{\ref{birmingham}} 
\and C.~Ziegler\inst{\ref{Dep_Phys_Stephen}} 
\and F.~Zong~Lang\inst{\ref{unibe}} 
\and S.~Z\'u\~niga-Fern\'andez\orcidZ{}\inst{\ref{astro_liege}} 
}

\institute{
Astrobiology Research Unit, Universit\'e de Li\`ege, All\'ee du 6 Ao\^ut 19C, B-4000 Li\`ege, Belgium \label{astro_liege}
\and Department of Astronomy \& Space Sciences, Faculty of Science, Ankara University, TR-06100, Ankara, T\"urkiye \label{Dep_turkey}
\and Ankara University, Astronomy and Space Sciences Research and Application Center (Kreiken Observatory), Incek Blvd., TR-06837, Ahlatlıbel, Ankara, T\"urkiye \label{Obs_turkey}
\and Ankara University, Graduate School of Natural and Applied Sciences, Department of Astronomy and Space Sciences, Ankara, T\"urkiye \label{institute_ankara}
\and Instituto de Astrof\'isica de Canarias (IAC), Calle V\'ia L\'actea s/n, 38200, La Laguna, Tenerife, Spain \label{IAC_Laguna} 
\and Department of Earth, Atmospheric and Planetary Science, Massachusetts Institute of Technology, 77 Massachusetts Avenue, Cambridge, MA 02139, USA \label{MIT}
\and Instituto de Astrof\'isica de Andaluc\'ia (IAA-CSIC), Glorieta de la Astronom\'ia s/n, 18008 Granada, Spain \label{iaa}
\and School of Physics \& Astronomy, University of Birmingham, Edgbaston, Birmingham B15 2TT, UK \label{birmingham}
\and Department of Physics and Kavli Institute for Astrophysics and Space Research, Massachusetts Institute of Technology, Cambridge, MA 02139, USA \label{Kavli_MIT}
\and Department of Astronomy \& Astrophysics, UC San Diego, La Jolla, CA 92093, USA \label{UCSD}
\and School of Physics, Faculty of Science, University of New South Wales, Kensington, Sydney, NSW 2052, Australia. \label{New_south_wales}
\and INAF- Palermo Astronomical Observatory, Piazza del Parlamento, 1, 90134 Palermo, Italy \label{INAF_Pa}
\and Center for Astrophysics \textbar \ Harvard \& Smithsonian, 60 Garden Street, Cambridge, MA 02138, USA \label{Harvard_USA}
\and Cadi Ayyad University, Oukaimeden Observatory, High Energy Physics, Astrophysics and  Geoscience Laboratory, Faculty of sciences Semlalia, Marrakech, Morocco \label{ouca}
\and Department of Physics and Astronomy, University of Kansas, Lawrence, KS 66045, USA \label{Univ_Kansas}
\and NASA Exoplanet Science Institute-Caltech/IPAC, Pasadena, CA 91125, USA \label{Caltech_IPAC}
\and Paris Region Fellow, Marie Sklodowska-Curie Action \label{Paris_Region}
\and AIM, CEA, CNRS, Universit\'e Paris-Saclay, Universit\'e de Paris, F-91191 Gif-sur-Yvette, France \label{cea}
\and Center for Space and Habitability, University of Bern, Gesellschaftsstrasse 6, 3012, Bern, Switzerland \label{unibe}
\and Department of Physics and McDonnell Center for the Space Sciences, Washington University, St. Louis, MO 63130, USA \label{st_louis}
\and NSF NOIRLab, 950 N. Cherry Ave., Tucson, AZ 85719, USA \label{NSF_NOIRLab}
\and Departamento de Astrof\'isica, Universidad de La Laguna (ULL), E-38206 La Laguna, Tenerife, Spain \label{Univ_LaLaguna}
\and Komaba Institute for Science, The University of Tokyo, 3-8-1 Komaba, Meguro, Tokyo 153-8902, Japan \label{Univ_tokyo}
\and Jet Propulsion Laboratory \label{Jet_Prop_Lab}
\and Grand Pra Observatory, 1984 Les Hauderes, Switzerland \label{Grand_Obs}
\and Universidad Nacional Aut\'onoma de M\'exico, Instituto de Astronom\'ia, AP 70-264, CDMX  04510, M\'exico \label{unamcu}
\and Department of Multi-Disciplinary Sciences, Graduate School of Arts and Sciences, The University of Tokyo, 3-8-1 Komaba, Meguro, Tokyo 153-8902, Japan \label{Dep_Multi_Dsci_Japan}
\and Okayama Observatory, Kyoto University, 3037-5 Honjo, Kamogatacho, Asakuchi, Okayama 719-0232, Japan \label{Okayama_Japan}
\and Cavendish Laboratory, JJ Thomson Avenue, Cambridge CB3 0HE, UK \label{Cavendish}
\and Space Sciences, Technologies and Astrophysics Research (STAR) Institute, Universit\'e de Li\`ege, All\'ee du 6 Ao\^ut 19C, B-4000 Li\`ege, Belgium \label{star_liege}
\and NASA Ames Research Center, Moffett Field, CA 94035, USA \label{Ames_NASA}
\and Department of Physical Sciences, Ritsumeikan University, Kusatsu, Shiga 525-8577, Japan \label{Ritsumeikan_Japan}
\and D\'epartement d'Astronomie, Universit\'e de Gen\'eve, Chemin Pegasi 51, CH-1290 Versoix, Switzerland \label{geneva}
\and Astrobiology Center, 2-21-1 Osawa, Mitaka, Tokyo 181-8588, Japan \label{Astro_tokyo}
\and Institute for Particle Physics and Astrophysics , ETH Z\"urich, Wolfgang-Pauli-Strasse 2, 8093 Z\"urich, Switzeland \label{ETH_Zur_Queloz}
\and Universidad Nacional Aut\'onoma de M\'exico, Instituto de Astronom\'ia, AP 106, Ensenada 22800, BC, M\'exico \label{uname}
\and Sternberg Astronomical Institute Lomonosov Moscow State University \label{Sternberg_Ast}
\and Lowell Observatory, 1400 W Mars Hill Road, Flagstaff, AZ, 86001, USA \label{Lowell_Obs}
\and Kotizarovci Observatory, Sarsoni 90, 51216 Viskovo, Croatia \label{Kotiza_Obs}
\and Department of Physics, Engineering and Astronomy, Stephen F. Austin State University, 1936 North St, Nacogdoches, TX 75962, USA \label{Dep_Phys_Stephen}
}

\date{Received/accepted}
\titlerunning{TOI-1743\,b, TOI-5799\,b, TOI-5799\,c and TOI-6223\,b}\authorrunning{S. Yal\c{c}{\i}nkaya et al.}

\abstract{
We present the discovery by the TESS mission of one transiting Neptune-sized planet, TOI-6223\,b, and two transiting super-Earths, TOI-1743\,b and TOI-5799\,b. We validate these planets using a statistical validation method, multi-color light curves, and other ancillary observations. We combined TESS and ground-based photometric data to constrain the physical properties of the planets. 

TOI-6223\,b is slightly larger than Neptune ($R_p=5.12^{+0.24}_{-0.25}$~$R_\oplus$) orbiting an early M dwarf in 3.86 days, and it has an equilibrium temperature of $T_{\rm eq}=714\pm14$~K. 
TOI-1743\,b orbits its mid-dwarf star  every 4.27~days. It has a radius of $R_p=1.83^{+0.11}_{-0.10}$~$R_\oplus$, and an equilibrium temperature of $T_{\rm eq}=485^{+14}_{-13}$~K. 
TOI-5799\,b has a radius of $R_p=1.733^{+0.096}_{-0.090}$ $R_\oplus$, and an equilibrium temperature of $T_{\rm eq}=505\pm16$~K orbit around an M2 dwarf in 4.17 days. 

We also present the discovery of an additional transiting planet, TOI-5799\,c, that we identified in the TESS data and validated using the SHERLOCK pipeline. TOI-5799\,c is a super-Earth with a radius of $R_p=1.76^{+0.11}_{-0.10}$~$R_\oplus$. Its orbital period and its equilibrium temperature are 14.01 days and  $T_{\rm eq}=337\pm11$~K, respectively, which place it near the inner edge of the habitable zone of its star.

We show that these planets are suitable for both radial velocity follow-up and atmospheric characterization. They orbit bright (< 11 $K_{mag}$) early M dwarfs, making them accessible for precise mass measurements. The combination of the planet sizes and  stellar brightness of their host stars also make them suitable targets for atmospheric exploration with the JWST. Such studies may provide insights into planet formation and evolution, as TOI-1743\,b, TOI-5799\,b, and TOI-5799\,c lie within the so-called radius valley, while TOI-6223\,b is located on the Neptunian ridge in the period--radius plane.}


\maketitle

\section{Introduction}

The bulk of exoplanet discoveries and their subsequent characterization have rapidly advanced in the past two decades, thanks to the prolific space missions Kepler \citep{Borucki2010} and the Transiting Exoplanet Survey Satellite (TESS, \citealt{Ricker2015}). Both of these missions aimed to detect exoplanet transits, but with a crucial distinction: Kepler focused on finding exoplanets around Sun-like stars in a specific region of the sky, while TESS is an all-sky survey designed to search for transits primarily around nearby low-mass stars, with each mission employing photometric filters tailored to their respective target populations. Kepler’s sensitivity to faint stars enabled the discovery of thousands of exoplanets, but it also renders most of them unsuitable for detailed follow-up characterization, particularly those with small radii relative to their host stars. Meanwhile, M dwarfs are the most abundant type of stars \citep{2010ARA&A..48..339B} and their smaller radii and masses make them favorable targets to search for exoplanets with the most fruitful techniques: transit \citep{2010exop.book...55W} and radial velocity (RV, \citealt{2016ASSL..428....3H}). Moreover, the habitable zone (HZ; \citealt{1993Icar..101..108K,2013ApJ...767L...8K}) of an M dwarf is much closer to the host star than that of earlier-type stars. As a result, both the transit and radial velocity methods, which favor the detection of close-in planets, are more likely to find planets within the HZ of M dwarfs (e.g. \citealt{Gillon_2016Natur,Gillon_2017Natur,Gilbert2020, Gilbert2023, 2022A&A...667A..59D}).

We report the validation of four exoplanet candidates identified in TESS data: TOI-1743\,b, TOI-5799\,b, TOI-5799\,c, and TOI-6223\,b. Three of these exoplanets (TOI-1743\,b, TOI-5799\,b and TOI-5799\,c) are particularly interesting because they lie within the so-called radius valley \citep{2017AJ....154..109F}—a region around $\sim$1.8$R_\oplus$ where there is a noticeable scarcity of sub-Neptune exoplanets with orbital periods shorter than 100 days around Sun-like stars. \citet{2020AJ....159..211C} confirmed the existence of the radius valley by applying completeness corrections, and also found that its central location shifts toward smaller planetary radii as the stellar mass decreases.

Several mechanisms have been proposed to explain the bimodal distribution observed in the radii of sub-Neptune (<4 $R_\oplus$) exoplanets, which likely reflects processes of planet 
formation and evolution. The left side of the distribution, which peaks at $\sim$1.3$R_\oplus$, may have been populated by small sub-Neptunes that formed during the later stages when gas was depleted in discs (gas poor formation; \citealt{2014ApJ...797...95L,2021ApJ...908...32L,2022ApJ...941..186L}). If these planets instead formed in protoplanetary disc with a rocky core smaller than 1.5 $R_\oplus$, surrounded by an envelope with a low mean molecular weight, their atmospheres could have been subject to photoevaporation (e.g. \citealt{2003ApJ...598L.121L, 2013ApJ...776....2L, 2013ApJ...775..105O, 2016ApJ...831..180C}). In addition to photoevaporation, radiation released during the cooling of rocky cores may also erode the atmosphere (core-powered mass loss; e.g. \citealt{2018MNRAS.476..759G, 2019MNRAS.487...24G}) or by giant accretionary collisions during the proto-planet formation stage (impact erosion; e.g., \citealt{2020ApJ...901L..31K}). 

\cite{2022Sci...377.1211L} proposed a density valley rather than a radius valley for exoplanets orbiting M-type stars, which aligns with the formation models that include orbital migration \citep{2018MNRAS.479L..81R}. 
The density valley separates sub-Neptune planets into three categories: rocky, water-rich, and gas-rich. Their analysis was based on 34 transiting sub-Neptune planets with precise mass measurements available at the time.
Distinguishing between water-rich and gas-rich sub-Neptunes based on density alone remains challenging, as interior and atmospheric composition models are degenerate. Therefore, expanding the sample size of sub-Neptunes suitable for both precise mass measurements and atmospheric characterization is essential for better understanding this distinction which can illuminate their formation and evolution history.

A similar radius distribution is also observed when considering the overall exoplanet population. A dearth of Neptune-sized planets is evident in close-in orbits, a feature widely recognized as the Neptunian desert or sub-Jovian desert  \citep{2011A&A...528A...2B, 2011ApJ...727L..44S, 2011ApJ...742...38Y,2013ApJ...763...12B,2016NatCo...711201L,2016A&A...589A..75M}. Photoevaporation is believed to be a primary mechanism shaping the boundaries of the Neptunian desert (e.g. \citealt{2019AREPS..47...67O}). At long period orbits relative to the Neptunian desert, the region becomes moderately populated and has been referred to as the Neptunian savanna by \cite{2023A&A...669A..63B}. A transition region between desert and savanna has recently been identified by \cite{2024A&A...689A.250C}. This region has an abrupt density of planets and proposed to be called as Neptunian ridge, where TOI-6223 b is placed.  \cite{2024A&A...689A.250C} suggest that the Neptunian ridge formed by exoplanets originally formed beyond snow line \citep{1996Icar..124...62P} and then arrived at their current position recently via high eccentricity migration (HEM) \citep{2003ApJ...589..605W, 2008ApJ...686..621F,2008ApJ...686..580C,2011CeMDA.111..105C,2012ApJ...751..119B}. In contrast, planets in the savanna may have reached these locations via disk-driven migration \citep{1979ApJ...233..857G,1996Natur.380..606L,2016SSRv..205...77B} and tend to have low orbital tilts and nearly circular orbits \citep{2023A&A...669A..63B}. Increasing the sample of exoplanets in these regions, suitable for precise radius and radial velocity follow-up observations to measure mass, eccentricity, and spin-orbit angle is crucial for understanding the characteristics, formation, and evolutionary history of Neptunian populations.

This paper is organized as follows. In Section \ref{sec:obs}, we describe the TESS and ground based observations. In Section \ref{sec:sherlock}, we describe our search for additional transiting exoplanets in the TESS light curves. In Section \ref{sec:validationmain}, we present the validation of exoplanet candidates detected by TESS. In Section \ref{sec:analysis} we describe the data analysis and present the stellar, orbital and absolute parameters of the systems. We discuss the planetary nature, future prospects and contribution to the exoplanet formation and evolution of exoplanets discovered within this study in Section \ref{sec:concanddiscs}. Finally, we present our conclusions in Section \ref{sec:conclusions}.

\section{Observations}\label{sec:obs}

\subsection{TESS photometric observation}

TIC~219860288 (TOI-1743) was observed with \emph{TESS} in sectors 15--26 (from Aug 15, 2019 to Jul 04, 2020), 40--41 (from Jan 24, 2021 to Aug 20, 2021), 47--56 (from Dec 30, 2021 to Sept 30, 2022), 58--60 (from Oct 29, 2022 to Jan 18, 2023), 73--83 (from Dec 07, 2023 to Oct 01, 2024) and 85--86 (from Oct 26 to Dec 18, 2024) with short and long cadences.
TIC~328081248 (TOI-5799) was observed in sectors 54 from Jul 9, 2022 to Aug 5, 2022, and 81 from Jul 15, 2024 to Aug 10, 2024 with 10-minute and 2-minute cadences.
TIC~288144647 (TOI-6223) was observed in sectors 57 from Sept 30, 2022 to Oct 29, 2022 with 200-second cadence and 84 from Oct 1 to 26, 2024 with 120-second and 200-second cadences.

Transits of TOI-6223.01 and TOI-5799.01 were detected by the Science Processing Operations Center \citep[SPOC,][]{SPOC_Jenkins_2016SPIE} pipeline and transit of TOI-1743.01 was detected by TESS Faint-star Search pipeline \citep{2022ApJS..259...33K}.

For our global analysis of the \emph{TESS} photometric data, we used the {\tt lightkurve} \citep{Lightkurve_2018ascl.soft12013L} Python package to download the presearch data conditioning light curves (PCD-SAP) extracted from the Mikulski Archive for Space Telescopes \citep{Stumpe_2012PASP,Smith_2012PASP,Stumpe_2014} constructed by the \emph{TESS} SPOC at Ames Research Center. PDC-SAP data have been corrected for any instrumental systematics and contamination from known nearby stars.

\begin{figure*}[!ht]
	\centering
	\includegraphics[width=0.65\columnwidth]{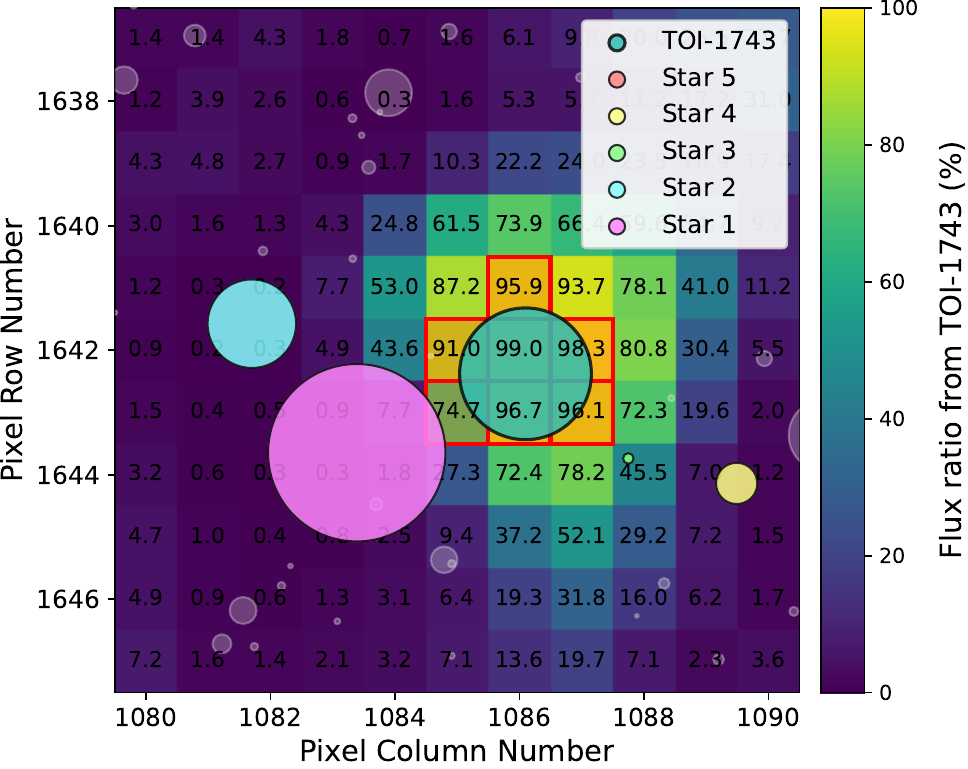} 
	\includegraphics[width=0.65\columnwidth]{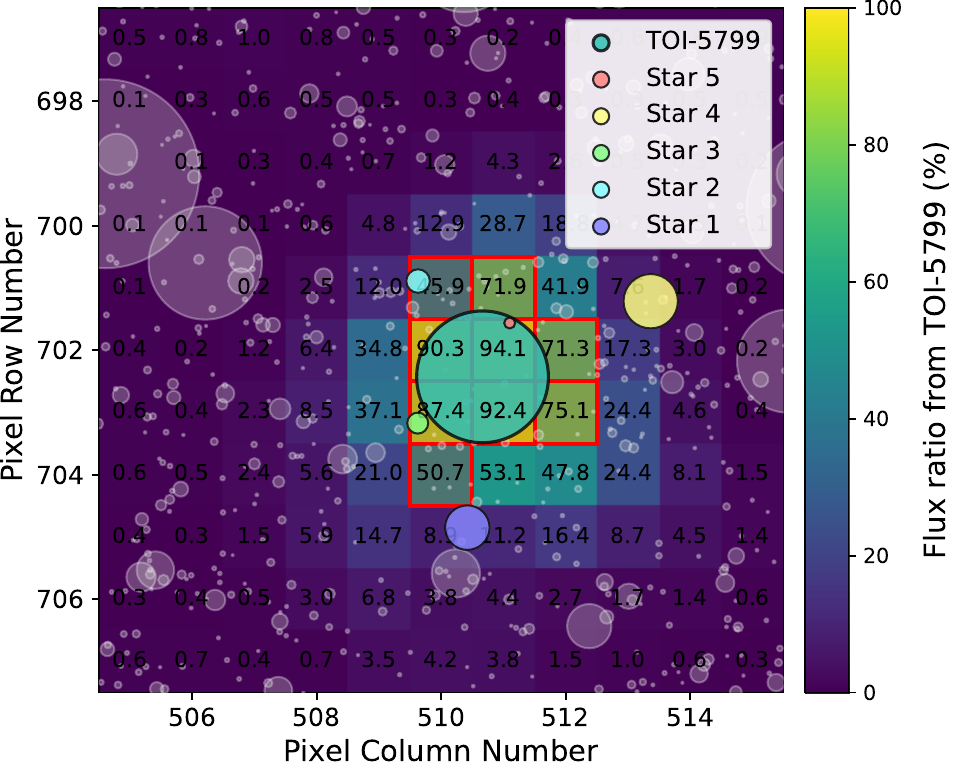} 
	\includegraphics[width=0.65\columnwidth]{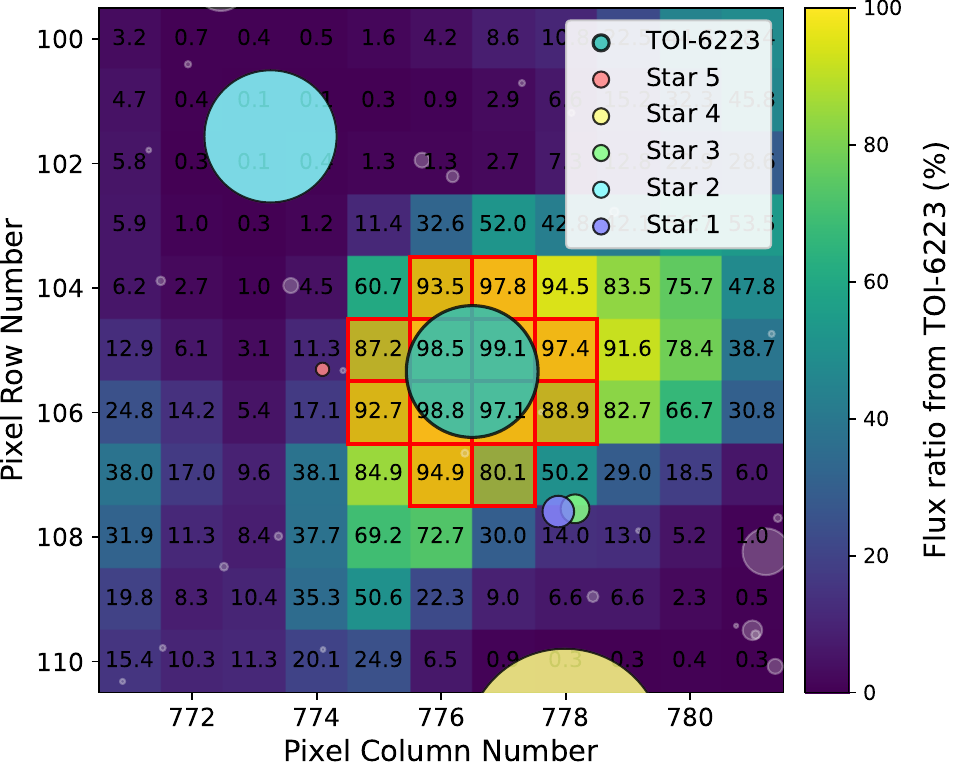}  
	\caption{Contaminating sources within the SPOC aperture (red mask) overlaid on the heat map from TESS observations of the target stars: TOI-1743 (left panel), TOI-5799 (middle panel) and TOI-6223 (right panel). The sizes of the target and nearby stars are scaled according to their relative fluxes in the TESS filter. The pixel scale is 21$\arcsec$ per pixel. This plot was generated using \texttt{TESS-cont} \citep{2024A&A...691A.233C}.}
	\label{fig:TESS_cont1}
\end{figure*}

\subsection{Ground-based photometric follow-up}\label{sec:gb}
After the detection of the transit signals by TESS, we observed these transits using ground-based telescopes to identify the source of the events within the seeing limits, improve the precision of timing and transit parameters, and examine the chromaticity of the transit depth. Information about these observations, including photometric filters, apertures, and observation dates, is listed in Table~\ref{obs_table}.

\subsubsection{TUG T100 observation}\label{sec:T100}
We observed a total of nine transits (two of TOI-1743.01, three of TOI-5799.01, one of TOI-5799.02, and three of TOI-6223.01) with the 1 meter Turkish telescope T100\footnote{https://tug.tubitak.gov.tr/en/teleskoplar/t100-telescope}, located at the Bak{\i}rl{\i}tepe / Antalya campus (TUG) of the T\"urkiye National Observatories at an altitude of 2500\,m. The telescope is equipped with a cryo-cooled SI 1100 CCD, which has $4096 \times 4096$ pixels and has a FoV of $21\arcmin \times 21\arcmin$. We used the $2 \times 2$ binning mode to reduce the read-out time to 15 seconds from 45 seconds in order to acquire more frames per observation. The data reduction and differential photometry relative to a synthetic star created by summing set of comparison stars was performed using the {\tt AstroImageJ} (AIJ) software \citep{collins2017}.

\subsubsection{AUKR T80 observation}
A partial transit of TOI-5799\,c was observed on UT October 23, 2024 in the Sloan-$i'$ filter with the 80cm Prof. Dr. Berahitdin Albayrak Telescope (T80, \citealt{2021AcA....71..223Y, 2023TJAA....4S.294O}), located at the Ankara University Kreiken Observatory (AUKR). The telescope is equipped with a $1024 \times 1024$ Apogee Alta U47+ CCD camera which has a FoV of $11\arcmin \times 11\arcmin$. We followed the same procedure described in the \ref{sec:T100} for data reduction and photometry. 

\begin{figure}[!ht]
    \centering
    \includegraphics[width=0.95\columnwidth, keepaspectratio]{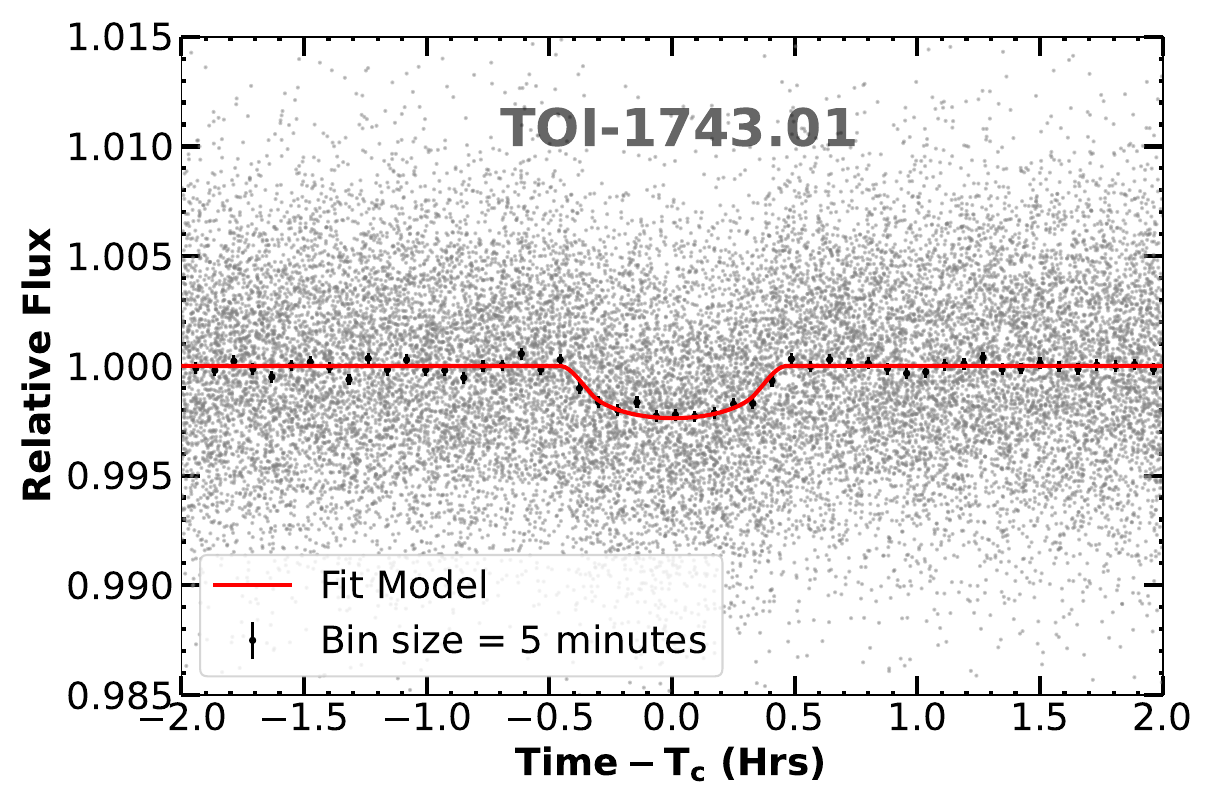}
    \includegraphics[width=0.95\columnwidth, keepaspectratio]{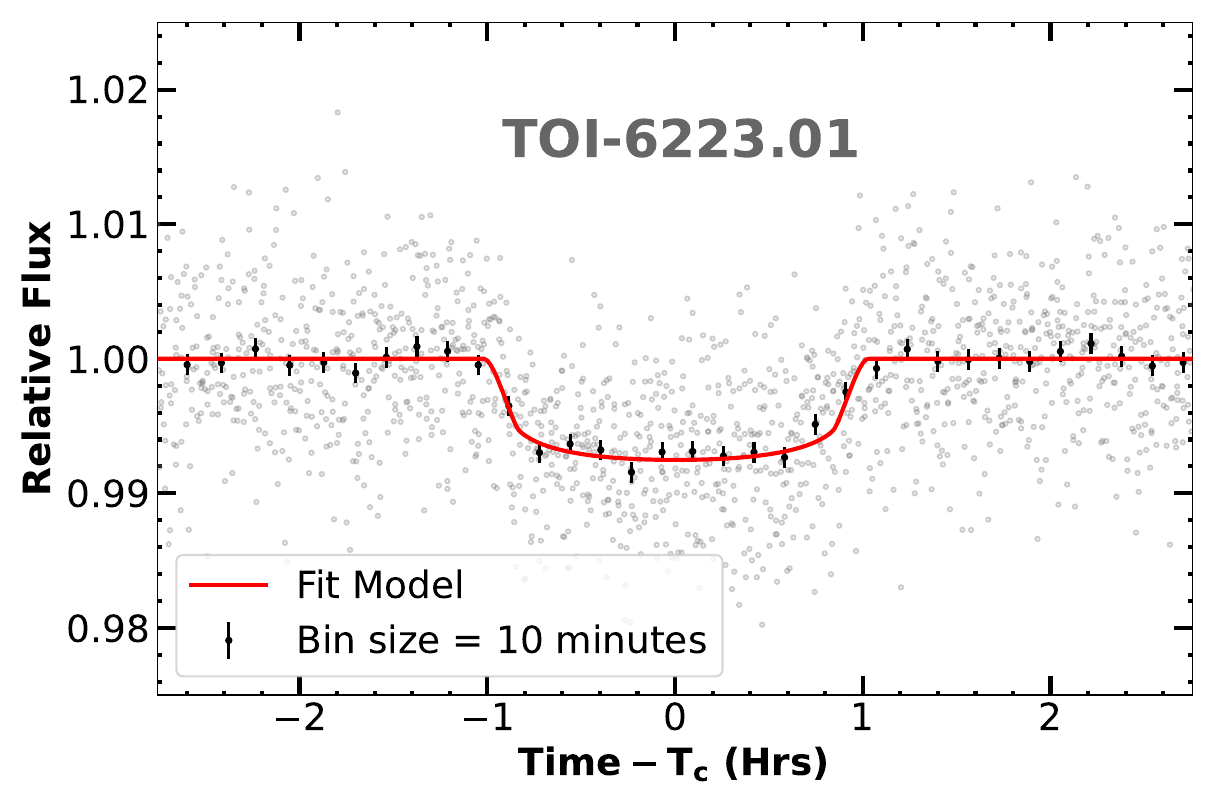}
    \caption{
        Phase-folded TESS transit light curves (gray circles) of TOI-1743.01 (\textit{top}) and TOI-6223.01 (\textit{bottom}). Global transit model (with limb darkening for TESS filter) for each system are shown with a red line. For clarity, binned light curves are shown with black dots.
}
    \label{fig:TESS1}
\end{figure}

\begin{figure}[!ht]
    \centering
    \includegraphics[width=0.95\columnwidth, keepaspectratio]{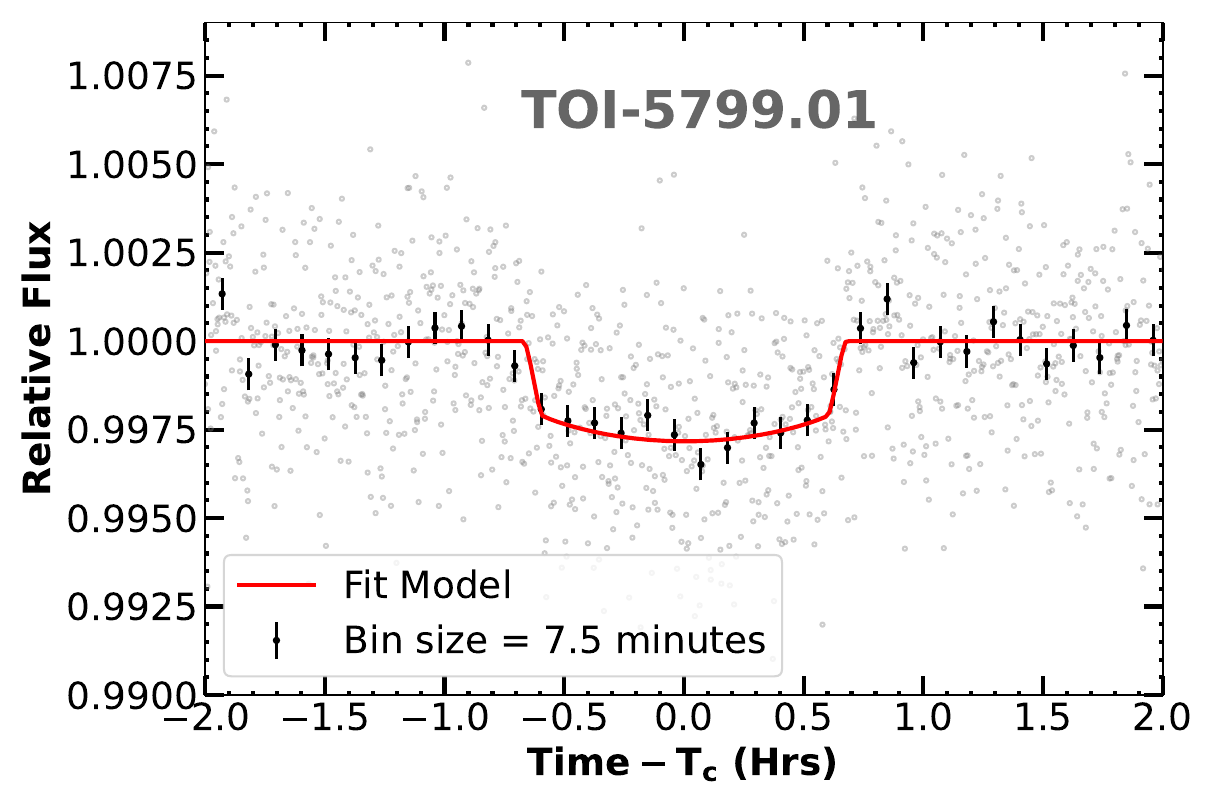}
    \includegraphics[width=0.95\columnwidth, keepaspectratio]{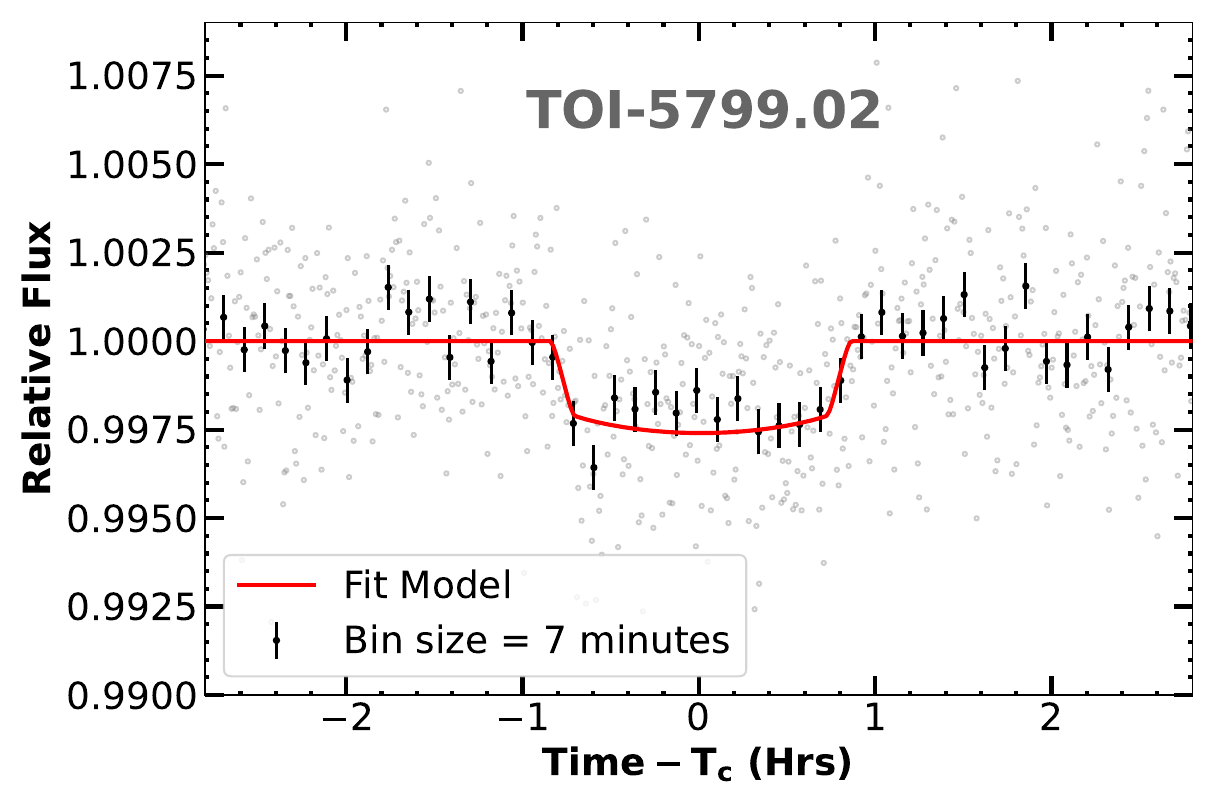}
    \caption{
        Same as Fig. \ref{fig:TESS1} but for inner (\textit{top}) and outer (\textit{bottom}) planets of TOI-5799 system.
}
    \label{fig:TESS2}
\end{figure}

\subsubsection{SPECULOOS-North and SAINT-EX observations}

SPECULOOS-North \citep[Search for habitable Planets EClipsing ULtra-cOOl Stars,][]{Burdanov2022} is a 1.0m telescope installed at Teide observatory since 2018. The telescope is equipped with a 2K$\times$2K Andor iKon-L BEX2-DD detector with a pixel scale of $0.35\arcsec$ and a FOV of $12\arcmin\times12\arcmin$. It is a twin of the  SPECULOOS-South  observatory  \citep{Jehin2018Msngr,Delrez2018,Sebastian_2021AA}  located at ESO-Paranal (Chile), and of the SAINT-EX observatory \citep{demory2020} located at the Sierra de San Pedro M\'artir (M\'exico).

SPECULOOS-North observed one full transit of TOI-5799.01 on UT October 25, 2024, while  two full transits of TOI-6223.01 on UT July 27 and UT October 20, 2024. All observations were conducted in the Sloan-$r'$ filter.
SAINT-EX observed a full transit of TOI-1743.01 on UT May 5, 2024 in the Sloan-$z'$ filter.
Science images processing and  photometric extraction were performed using the {\tt PROSE}\footnote{{\tt Prose:} \url{https://github.com/lgrcia/prose}} pipeline \citep{prose}. 

\subsubsection{TRAPPIST-North-0.6m observation}

We observed two full transits of TOI-1743.01 and one transit of TOI-5799.01 with the TRAPPIST-North telescope. 
TRAPPIST-North \citep{Barkaoui2019_TN} is a 0.6m Ritchey-Chr\'etien robotic telescope
in operation at Oukaimeden Observatory since June 2016. It is equipped with a thermoelectrically cooled 2K$\times$2K Andor iKon-L BEX2-DD CCD camera with a pixel scale of 0.6\arcsec\, resulting a FOV of $20\arcmin\times20\arcmin$. TRAPPIST-North is a twin of the TRAPPIST-South \citep{Gillon2011,Jehin2011} located at La Silla Observatory in Chile since 2010. 
TOI-1743.01 observations were carried out on UT July 19 and September 21, 2020 in the $I+z$ filter, while TOI-5799.01 observation was carried out on UTC September 28, 2022 in the Sloan-$z'$ filter. 
Science images processing and photometric extraction were performed using the {\tt PROSE} pipeline. 

\subsubsection{LCO-HAL-2m0/MuSCAT3 observation}\label{sec:muscat3}

LCO-HAL-2m0 observed a full transit of TOI-1743.01 on UT May 27, 2021.
It is located at Haleakala Observatory Maui, Hawaii and equipped with MuSCAT3 multi-band imager \citep{Narita_2020SPIE11447E}.
The observations were conducted simultaneously in the Sloan-$g'$, -$r'$, -$i'$, and Pan-STARRS-$z$ filters.
Science images calibration and photometric extraction were performed using the LCOGT {\tt BANZAI} pipeline \citep{McCully_2018SPIE10707E} and {\tt AIJ}, respectively.

\subsubsection{MuSCAT2 observation}
We observed one full transit of TOI-5799.01 and two full transits of TOI-6223.01 with the MuSCAT2 multicolor imager \citep{Narita2018} mounted on the 1.52m-Telescopio Carlos S\'anchez (TCS) at the Teide Observatory in Tenerife (Canary Islands, Spain).
All the transit observations were carried out simultaneously in the Sloan-$g$, -$r$, -$i$, and $z_\mathrm{s}$.
TOI-5799.01 was observed on UT July 8, 2023; while TOI-6223.01 was observed on UT August 23 and September 23, 2024. Data reduction and photometry were carried out using the MuSCAT2 photometry pipeline \citep{Parviainen2020}.

\subsubsection{LCOGT-1m0 observation}
Full transits of TOI-1743.01 on UT July 7, 2020 and TOI-5799.01 on UT June 14, 2023 observed using 1 meter the Las Cumbres Observatory Global Telescope (LCOGT; \citealt{Brown_2013}) network. The telescopes are equipped with a $4096\times4096$ SINISTRO detectors, having a pixel scale of $0\farcs389$ and FOV of $26\arcmin\times26\arcmin$.

TOI-1743.01 was observed with LCO-McD-1.0m at McDonald Observatory in the USA, while TOI-5799.01 was observed at LCO-Teid-1.0m at Teide Observatory in Tenerife (Spain). Both observations were conducted in the Sloan-$i'$ filter. Science images calibration and photometric extraction were performed using the LCOGT {\tt BANZAI} pipeline and {\tt AIJ}, respectively. 

\subsubsection{KeplerCam observation}
We used KeplerCam CCD to observe a full transit of TOI-1743.01 on UT April 10, 2021. The KeplerCam is a $4096\times4096$ pixel CCD camera mounted on the 1.2-m telescope located at the Fred Lawrence Whipple Observatory (FLWO) in Arizona (USA). The observations were carried out in the Sloan-$i'$ filter with 2$\times$2 binning mode, resulting in a pixel scale of 0.672\arcsec/pixel. 
We followed the same procedure described in the \ref{sec:T100} for data reduction and photometric extraction. 

\subsubsection{RCO-0.4m observation}

The detection of transits of TOI-1743.01 was obtained on UT June 02, 2020 and on UT September 04, 2020 with an FLI 4710 camera mounted on a RCO 40 cm telescope located at the Grand-Pra Observatory, Switzerland. FLI4710 is an $11.7\arcmin\times11.7\arcmin$ FoV back illuminated CCD using an E2V CCD47-10 sensor. Observations were taken in 1$\times$1 binning mode and produced a $0.73\arcsec$ per pixel scale. Observations were performed using Sloan-$i^{\prime}$ filter. We followed the same procedure described in the \ref{sec:T100} for data reduction and photometry. We detected an egress on the first observation and a full transit on the second, using uncontaminated 3.6 and 5.1 arcsecond-apertures, respectively.

\subsection{High-resolution Observations}
In order to detect the possible dilution of background/foreground or bound companions on the planetary radii derived from transit light curves \citep{ciardi2015}, the TOIs were observed with optical speckle and Robo-AO observations and near-infrared adaptive optics imaging.

\subsubsection{2.5m-SAI high-resolution observations}
\label{SAI_images}

TOI-6223 was observed on UT September 29, 2023, using the speckle polarimeter mounted on the 2.5-meter telescope at the Caucasian Observatory of the Sternberg Astronomical Institute (SAI), Lomonosov Moscow State University. The observations employed a low-noise Hamamatsu ORCA–quest CMOS detector \citep{Strakhov_2023AstBu}. An atmospheric dispersion compensator enabled the use of the $I_\mathrm{c}$ band, providing an angular resolution of $0.083\arcsec$. No companions were detected for target. The detection limits at distances $0.25$ and $1.0\arcsec$ from the star is $\Delta I_\mathrm{c}=4.6^m$ and $6.4^m$. The image and its contrast curve is shown in Figure \ref{fig:high_res_SAI}.

\subsubsection{Palomar high-resolution observations}\label{sec:palomar}

Observations of TOI-5799 were made on UT June 23, 2023 with the PHARO instrument \citep{hayward2001} on the Palomar Hale (5m) behind the P3K natural guide star AO system \citep{dekany2013}, and observations of TOI-1743 and TOI-6223 were made on UT April 20, 2020 and May 05, 2023 with the NIRC2 instrument on Keck-II (10m) behind the natural guide star AO system \citep{wizinowich2000}.
The pixel scales for PHARO and NIRC2 are $0.025\arcsec$ and $0.009942\arcsec$ per pixel. The Palomar data were collected in a standard 5-point quincunx dither pattern and the Keck data were collected in a 3-point dither pattern that avoids the lower left quadrant of the array. The calibrated data were combined into a single mosaic frame, resulting in a final resolution of $\sim 0.1\arcsec$ and $0.05\arcsec$ for PHARO and NIRC2, respectively.  
	
To estimate the detection sensitivity in the final combined AO images, we injected artificial sources azimuthally around the target star every $20^\circ$ positioned at separations corresponding to integer multiples of the FWHM of the central source \citep{furlan2017}. The brightness of each injected source was adjusted until it reached a $5\sigma$ detection threshold based on standard aperture photometry. At each separation, the final $5\sigma$ contrast limit was taken as the mean detection threshold across all position angles, with the uncertainty estimated from the RMS variation among those measurements. No nearby background objects are identified within our detection limits.

\subsubsection{Near-Infrared AO Imaging}
In order to further exclude potential contamination from nearby bound or line-of-sight companions, we observed TOI-6223 and TOI-1743 with NIRC2 adaptive optics (AO) imaging at Keck Observatory. TOI-6223 was observed on UT August 05, 2023 in the narrow band K continuum (\texttt{Kcont}; $\lambda_o = 2.2706; \Delta\lambda = 0.0296~\mu$m) filter. TOI-1743 was observed on UT May 28, 2022 in the wider band K filter (\texttt{K}; $\lambda_o = 2.196; \Delta\lambda = 0.336~\mu$m). Observations were taken behind the natural guide star systems \citep{dekany2013,wizinowich2000} with the narrow angle mode, providing a pixel scale of around $0.01\arcsec$ per pixel and a full FoV of about $10\arcsec$. To avoid the lower-left quadrant of the detector, which is usually noisier than the other quadrants, a standard three-point dither pattern with a step size of $3\arcsec$ was applied. Each dither position was imaged three times, with $0.5\arcsec$ positional offsets between exposures, resulting in a total of nine frames. The reduced science frames were combined into a single mosaic image with a final resolution of $0.05\arcsec$ and $0.06\arcsec$ for TOI-6223 and TOI-1743, respectively.

 We followed the same procedure described in Sec. \ref{sec:palomar} to determine sensitivity of the final combined images. The final images and sensitivities are shown in Figure~\ref{fig:high_res_Keck}. No other nearby stellar companions are identified within our detection limits.

\subsubsection{WIYN observations}
We observed TOI-1743 on UT April 20, 2021 using the NN-EXPLORE Exoplanet Stellar Speckle Imager (NESSI; \citealt{Scott_2018PASP}) with the WIYN 3.5~m telescope on Kitt Peak. NESSI can obtain simultaneous speckle images in two filters.  We used filters with central wavelengths $\lambda_c = 562$ and 832~nm for these data. The observation consisted of a set of 13000 40~ms exposures in each filter.  NESSI's field-of-view was restricted to $4.6\arcsec\times\arcsec4.6$ by reading out a $256\times256$ pixel sub-array of NESSI's EMCCD detectors centered on the target star.  Additionally, we confined photometric measurements to an outer radius of 1.2~arcseconds from the target star.  Along with TOI-1743, we observed a point-source PSF calibrator star. The calibration observation consisted of a single 1000-frame image set and is used to set the PSF of the speckle image reconstruction.

The NESSI data were processed using the pipeline described by \citet{Howell_2011AJ}. The Pipeline produces reconstructed image of the field around TOI-1743 for each filter and this image was used to derive contrast curves. The contrast limit corresponds to the level of fluctuations in the noise-like background of the reconstructed image as a function of distance from the target star. No companion sources were identified near TOI-1743 in the NESSI observations. Final images and the contrast curves in each filter are shown in Figure~\ref{fig:high_res_WIYN}).

\subsubsection{SOAR observations}
We observed TOI-5799 with the speckle imaging using 4.1-m Southern Astrophysical Research (SOAR) telescope \citep{Tokovinin_2018PASP} on UT November 04, 2022 in the Cousins I-filter. The $5\sigma$ detection limit was $\Delta$mag=3.2 at an angular distance of $1\arcsec$ from the target. Detailed information about data reduction, creation of contrast curve and SOAR observations of TESS candidates can be found in \citet{Ziegler_2020AJ}. No nearby sources were detected within $3\arcsec$ of TOI-5799 in the SOAR observations. Figure \ref{fig:high_res_WIYN} presents high resolution image, the $5\sigma$ detection limits and the speckle auto-correlation functions derived from the observation.

\subsection{Spectral Observations}

\subsubsection{Shane/KAST}\label{sec:Shane}
All three host stars were observed with the Kast Double Spectrograph \citep{kastspectrograph} on the 3m Shane Telescope at Lick Observatory. 
TOI-1743 was observed on UT April 28, 2024 in partly cloudy conditions with 2$\arcsec$ seeing; TOI-5799 and TOI-6223 were observed on UT 2024 December 06, 2024 in clear and windy conditions with 1.3$\arcsec$ and 1.8$\arcsec$ seeing.
Both the blue and red channels of Kast were used, with light split at 5700~{\AA} 
by the D57 dichroic.
Slit widths of 1$\arcsec$--2$\arcsec$ were used depending on seeing conditions.
For the blue channel, we used the 600/4310 grism which provides 3600--5600~{\AA} wavelength coverage at an average resolution of $\lambda/\Delta\lambda$ = 1470 for a slit width of 1$\arcsec$.
For the red channel, we used the 600/7500 grating which provides 5900--9000~{\AA} wavelength coverage at an average resolution of $\lambda/\Delta\lambda$ = 1960 for a slit width of 1$\arcsec$. Total exposures ranged from 300~s (TOI-5938) to 900~s (TOI-1743), split into a single blue and two sequential red exposures, and all sources were observed at an airmass $<$1.3.
We observed a nearby G2~V star for calibration of telluric absorption, and the spectrophotomeric flux standards Feige 66 (April 2024) Hiltner 600 (December 2024; \citealt{1992PASP..104..533H,1994PASP..106..566H}) were observed during each night for flux calibration.
For each night, the wavelength calibration and pixel response were performed using HgCd arc, HeNeAr arc, and  flat-field lamps.
Data reduction was performed with using the kastredux package\footnote{kastredux:~\url{https://github.com/aburgasser/kastredux}.}, which included image reduction, boxcar extraction of the 1D spectrum with median background subtraction, wavelength calibration, flux calibration, and correction of telluric absorption.
The final spectra have median signals-to-noise of 55-65 in the blue channel (near 5425~{\AA}) and 100-190 in the red channel (near 7450~{\AA}).

\begin{figure}
    \centering
    \includegraphics[width=0.85\columnwidth, keepaspectratio]{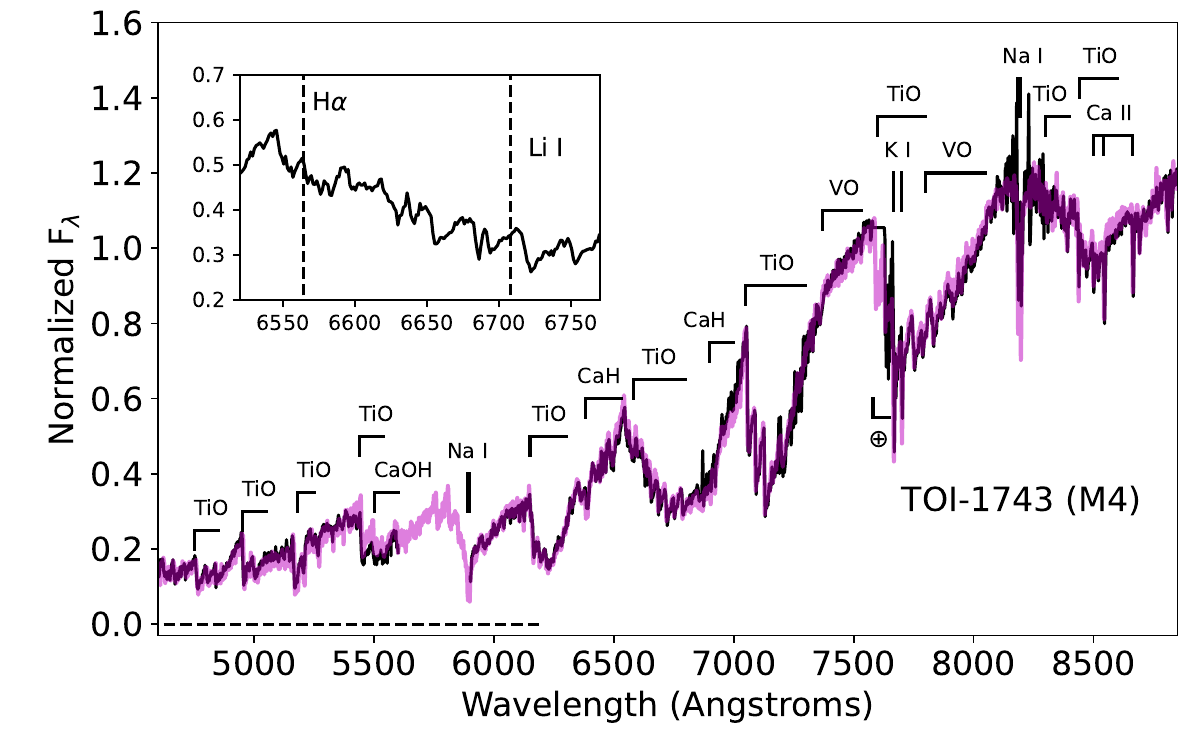} \\
    \includegraphics[width=0.85\columnwidth, keepaspectratio]{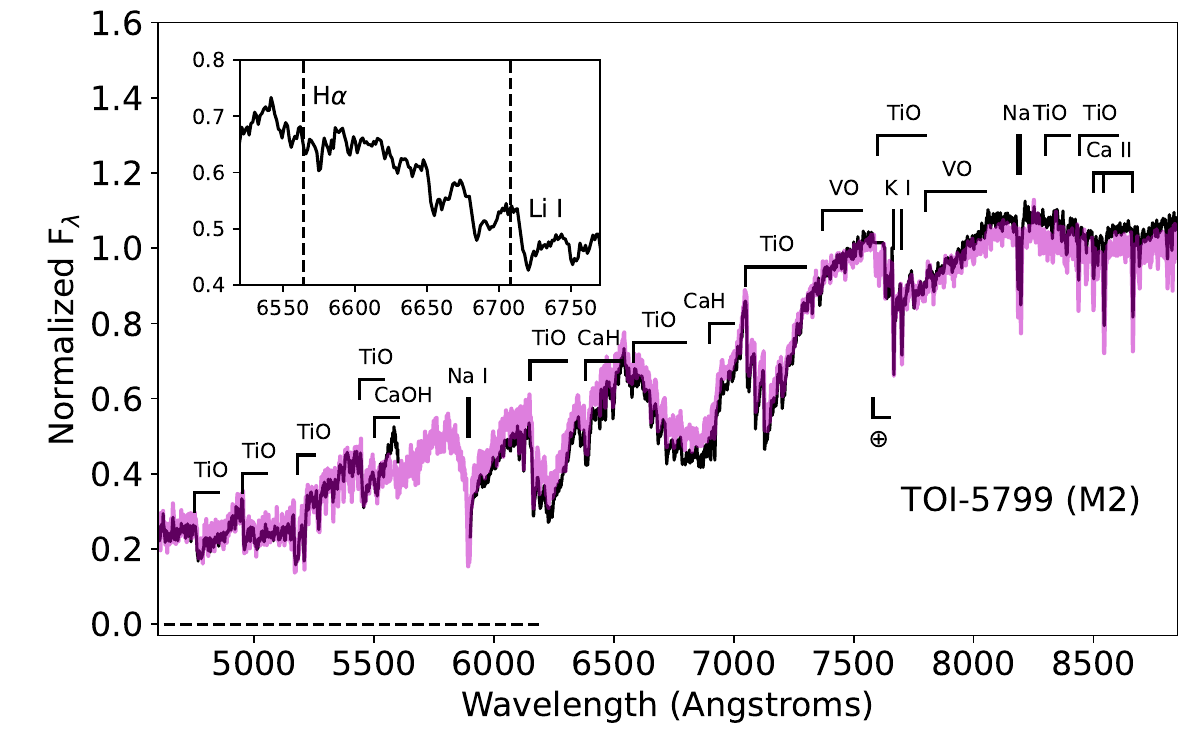} \\
    \includegraphics[width=0.85\columnwidth, keepaspectratio]{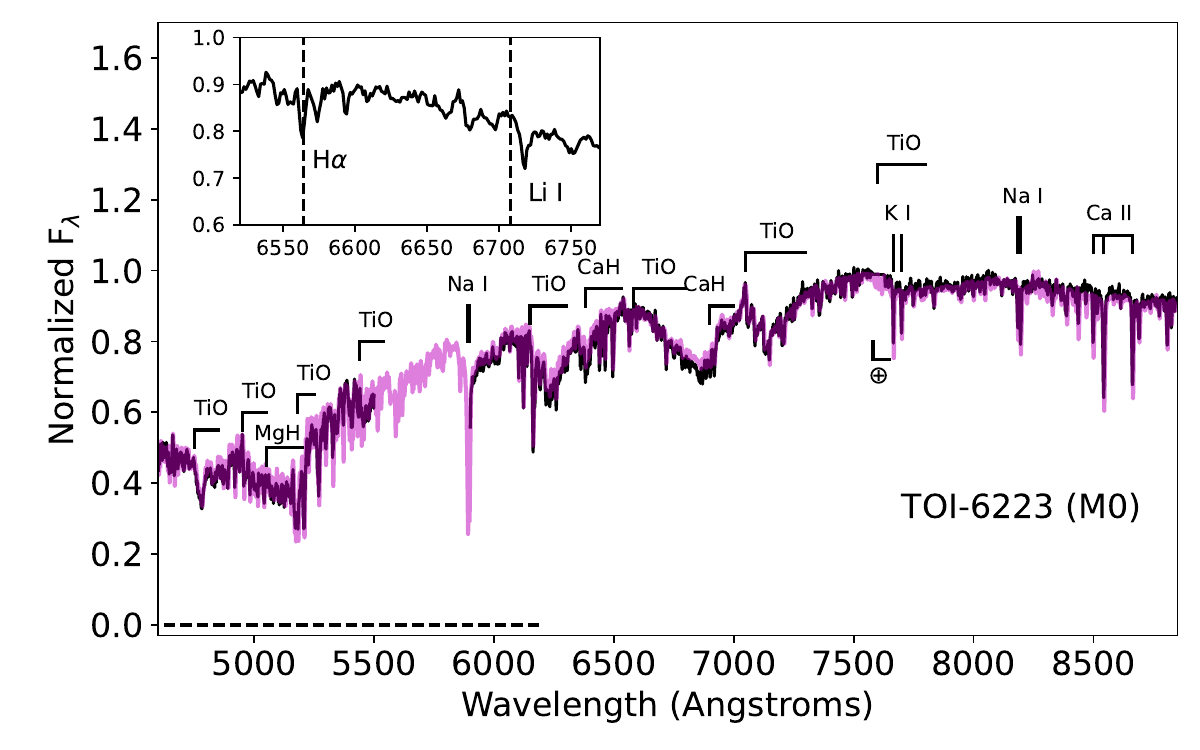}
    \caption{
        Black lines show the normalized (at 7400~{\AA}) Shane/Kast optical spectra of (from top to bottom): 
        TOI-1743, TOI-5799, and TOI-6223. Best-fit SDSS standard templates \citep{2007AJ....133..531B} are shown with magenta lines.
        The key spectral features between 4200--8900~{\AA} are labeled, including the location of the (corrected) O$_2$ telluric A-band at 7500~{\AA} ($\oplus$).
        Zoomed regions around  6563~{\AA} H$\alpha$ and 6708~{\AA} Li~I lines are shown in inset boxes.
        The 5600–5900~{\AA} gap due to the dichroic separation between the blue and red channels of the Kast spectrograph.       
}
    \label{fig:kast}
\end{figure}

Figure\,\ref{fig:kast} displays the reduced Shane/Kast spectra of the three targets compared to SDSS K and M dwarf templates from \citet{2007AJ....133..531B}.
From early-type to late-type, we infer classifications of
M0 for TOI-6223,
M2 for TOI-5799, and
M4 for TOI-1743.
These template classifications are further supported by index-based classifications from \citet{1995AJ....110.1838R}; \citet{1997AJ....113..806G}; and \citet{2003AJ....125.1598L}; and in the case of TOI-5799 our classification is identical to that reported by \citet{2015A&A...577A.128A}.
H$\alpha$ is seen in absorption in 
TOI-6223 (EW = 0.46$\pm$0.05~{\AA}),
in emission in TOI-1743 (EW = $-$0.83$\pm$0.09),
and no evidence of absorption or emission is seen in TOI-5799.
The presence of H$\alpha$ emission in TOI-1743 is consistent
with a relative luminosity of $\log_{10}{L_{H\alpha}/L_{bol}}$ = $-$4.53$\pm$0.08 and indicates an 
age $\lesssim$4.5~Gyr \citep{2008AJ....135..785W}. Stellar mass -- activity lifetime relation by \cite{2024ApJ...966..231P} suggests an even younger age ($\lesssim$0.7~Gyr) for TOI-1743.
Three of the sources have metallicity index values of $\zeta \approx 1$ \citep{2007ApJ...669.1235L,2013AJ....145...52M}, indicating approximately solar metallicities. The exception is TOI-6223 for which we measure $\zeta$ = 1.30$\pm$0.02 suggesting a significantly supersolar metallicity ([Fe/H] $\gtrsim$ $+0.2$;  \citealt{2013AJ....145...52M}).
We note that Gaia DR3 spectral measurements of this source indicate a slightly supersolar metallicity ([Fe/H] = +0.23$_{-0.08}^{+0.02}$ \citealt{2023A&A...674A..27A}) while analysis of LAMOST spectral observations indicate a slightly subsolar metallicity ([Fe/H] = $-$0.20$\pm$0.16 \citealt{2022ApJS..260...45D})

\subsubsection{Magellan/MagE}\label{sec:Magellan}

We observed TOI-5799 on UT May 22, 2023 with the Magellan Echellette spectrograph (MagE; \citealt{2008SPIE.7014E.169M}) mounted on the 6.5-m Magellan Baade Telescope. The weather conditions were clear and stable  and a seeing of $1\farcs4$. We used the 0.7\arcsec$\times$10$\arcsec$ slit to obtain a resolution of $\lambda/\Delta\lambda$ $\approx$ 6000 over 4000--9\,000\,{\AA}. We collected two 40-s exposures at an airmass of 1.50. For the flux calibration, we observed the spectrophotometric calibrator EG\,274 \citep{1992PASP..104..533H,1994PASP..106..566H} same night. 
For each night, the wavelength calibration and pixel response were performed using ThAr arc, Xe flash lamps, bias, and incandescent lamp images.
We did not obtain a telluric calibrator for these observations, therefore, strong atmospheric absorption features from oxygen and water remain visible in the spectrum. Data reduction was performed using PypeIt \citep{pypeit:zenodo, pypeit:joss_arXiv, pypeit:joss_pub} in standard settings. The reduced final spectrum has a median S/N $\approx$ 50 at 5450~{\AA} and S/N $\approx$ 100 at 7450~{\AA}. 

\begin{figure}
    \centering
    \includegraphics[width=1.0\columnwidth, keepaspectratio]{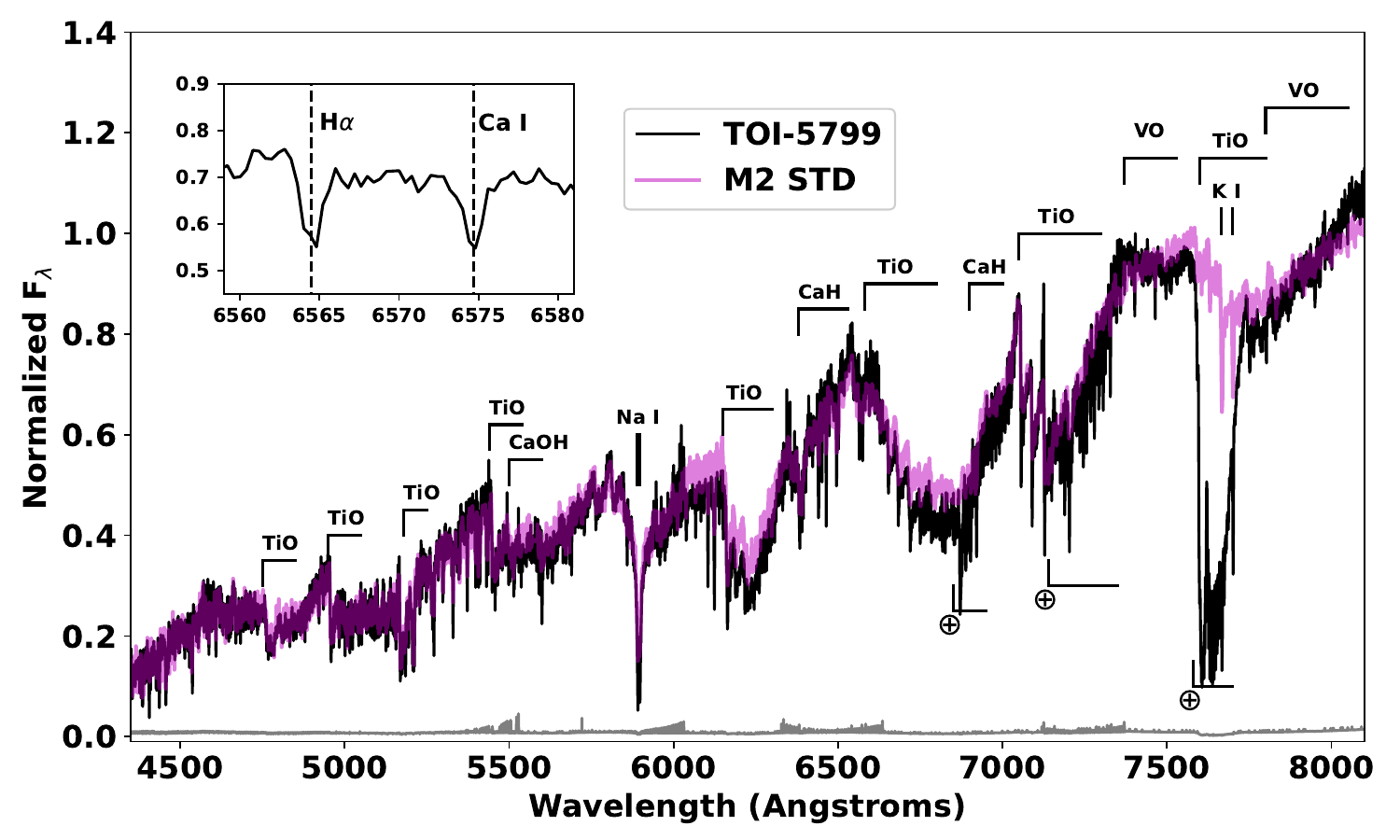}
    \caption{
        The Baade/MagE spectrum of TOI-5799 (black line) is shown in comparison with the M2 SDSS template from \citet[][; magenta line]{2007AJ....133..531B}. The spectra are normalized at 7400~{\AA}, and prominent spectral features between 4300--8100~{\AA} are labeled, including regions affected by uncorrected telluric oxygen and water absorption ($\oplus$). 6555–6585~{\AA} region is shown in inset box to highlight H$\alpha$ and Ca~I features.     
}
    \label{fig:mage}
\end{figure}

Figure\,\ref{fig:mage} displays the reduced MagE spectrum over the 4200--8100~{\AA} range. We conducted a similar analysis as our Kast observations, finding again a best-match SDSS spectral template of M2 that is confirmed by index-based classifications. The higher-resolution MagE data are able to discern a clear absorption feature from H$\alpha$ with EW = 0.30$\pm$0.03~${\AA}$, which places a relatively weak constraint on the age of the system ($\gtrsim$1.2~Gyr; \citealt{2008AJ....135..785W}). We measure a low $\zeta$ = 0.86$\pm$0.02 from this spectrum, consistent with a subsolar metallicity ([Fe/H] = $-$0.17$\pm$0.20 \citealt{2013AJ....145...52M}; $-$0.28$\pm$0.30 \citealt{2009PASP..121..117W}). Both \citet{2014AJ....147...20N} and \citet{2015ApJS..220...16T} have previously reported subsolar metallicities for this source based on near-infrared spectroscopy ([Fe/H] = $-$0.20$\pm$0.12, $-$0.30$\pm$0.11), while \citet{2022ApJ...927..122H} report a near-solar metallicity from optical spectroscopy ([Fe/H] = +0.05$\pm$0.22). The bulk of evidence suggests that this is a slightly metal-poor star.

\subsubsection{IRTF/SpeX}\label{sec:IRTF}

\begin{figure}
    \centering
    \includegraphics[width=\linewidth]{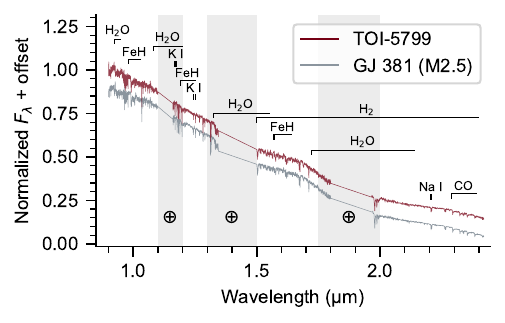}
    \includegraphics[width=\linewidth]{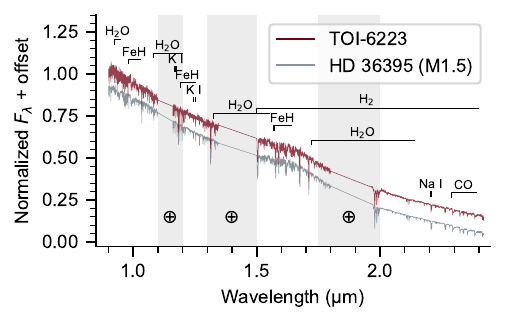}
    \caption{
    IRTF/SpeX SXD spectra of TOI-5799 (top) and TOI-6223 (bottom).
    The target spectra (red) are plotted alongside M-dwarf spectral templates (grey) and offset vertically for clarity.
    Prominent M-dwarf spectral features are labeled, and wavelengths strongly affected by telluric absorption are shaded.
    }
    \label{fig:spex}
\end{figure}

We obtained medium-resolution near-infrared spectra of TOI-5799 and TOI-6223 with the 3.2-m NASA Infrared Telescope Facility (IRTF) using the SpeX spectrograph \citep{Rayner2003}.
We used the short-wavelength cross-dispersed (SXD) mode with the 0$\farcs3$ $\times$ 15’’ slit aligned to the parallactic angle and nodded in an ABBAAB pattern.
TOI-5799 was observed on 18 May 2025 under clear conditions with 0$\farcs$9 seeing.
We collected six 200\,s exposures at an airmass of 1.0, followed by a standard SXD calibration set and six 20\,s exposures of the A0\,V telluric standard HD\,192538 ($V=6.5$) at a similar airmass.
TOI-6223 was observed on 19 June 2025 in 1$\farcs$1 seeing.
We gathered six 250\,s exposures at an airmass of 1.1, followed by a calibration set and six 15\,s exposures of HD\,209932 ($V=6.5$) at a similar airmass.
We reduced both datasets using Spextool v4.1 \citep{Cushing2004}, following standard procedures \citep[e.g.,][]{Ghachoui2023, Ghachoui2024, 2024A&A...687A.264B, Barkaoui2025_TOI-6508b}.
The resulting spectra (Figure\,\ref{fig:spex}) span 0.90–2.42\,$\mu$m at a resolving power of $R{\sim}2000$ with 2.5\,pixels per resolution element and have median per-pixel signal-to-noise ratios of 215 and 107, respectively

We used the SpeX SXD spectra and the SpeX Prism Library Analysis Toolkit \citep[SPLAT, ][]{splat} to estimate spectral types and metallicities for TOI-5799 and TOI-6223.
For TOI-5799, comparison to standards in the IRTF Spectral Library \citep{Cushing2005, Rayner2009} reveals a close match to the M2.5\,V standard GJ 381 (Figure\,\ref{fig:spex}), and we adopt a near-infrared spectral type of M2.5$\pm$0.5.
From $K$-band Na\,\textsc{i} and Ca\,\textsc{i} features and H2O–K2 index \citep{Rojas-Ayala2012} and using the \citet{Mann2013} relation, we estimate a metallicity of $\mathrm{[Fe/H]} = -0.35 \pm 11$, suggestive of a sub-solar metallicity.
For TOI-6223, we find a best match to the M1.5\,V standard HD\,36395 and adopt a near-infrared spectral type of M1.5$\pm$0.5.
Using the same features and empirical relation, we estimate $\mathrm{[Fe/H]} = +0.15 \pm 0.12$ for TOI-6223, consistent with solar values.

\subsubsection{TRES}
TRES  is a high-resolution spectrograph attached to the 1.5m Tillinghast Reflector telescope located at the Fred Lawrence Whipple Observatory (FLWO) in Arizona.
TRES is a fiber-fed spectrograph that observes in the 390--910\,nm range with a resolving power of R=44,000. Spectra were extracted using the methods described in \citet{buchhave2010}.
Three reconnaissance spectra of TOI-5799 were obtained on UT April 10, May 15 and 17, 2024.
Three reconnaissance spectra of TOI-6223 were obtained on UT Jan 2, 3 and June 18, 2024.
We extracted radial velocity measurements of TOI-5799 using a pipeline optimized for M-dwarf stars \citep{Pass_2023}, while TOI-6223 RVs extraction was performed using the TRES pipeline. We followed the same procedure as described in \citet{2024A&A...687A.264B}. 
H$\alpha$ appears in absorption and there is no detectable rotational broadening given the resolution of the spectrograph ($v\sin(i) < 3.4km/s$). The RVs rule out a brown dwarf or stellar transiting companion.

\section{Search For Additional Transiting Exoplanets}\label{sec:sherlock}

We analyzed publicly available TESS data to independently recover the officially alerted candidates and search for hints of potential ones that may have remained unnoticed by the SPOC and Quick-Look Pipeline (QLP) pipelines. 
To this end, we employed the \texttt{SHERLOCK} package \citep{pozuelos2020,demory2020} by combining all available sectors for each TOI and explored orbital periods from 1 to 30\,d, using ten detrended scenarios spanning from 0.2 to 1.2\,d size. We refer the reader to \cite{pozuelos2023} for further details on different search strategies and to \cite{devora2024} for the details of the most up-to-date version of the pipeline, including a full description of all its functionalities. 

In all the cases, we successfully recovered the alerted candidates, allowing us to confirm these alerts independently. On the one hand, for TOI-1743 and TOI-6223, we did not identify any additional signals that could be considered hints of planetary candidates. On the other hand, in the case of TOI-5799, we found a persistent secondary signal across all detrends applied to the data, with an orbital period of $\sim$14.02\,d, which caught our attention. We then used the vetting module included in \texttt{SHERLOCK} to evaluate the transit shape, odd-even consistency, centroid shifts, optical ghost effects, transit source offsets, rolling band contamination, and other factors. We refer the reader to \cite{devora2024} for further details. We did not find any potential warning for any of the parameters considered in this vetting. We then triggered the ground-based follow-up campaign to confirm the event in the target star and performed the statistical validation as described in Sec.~\ref{sec:gb} and \ref{sec:validation}, respectively. These analyses, although still lacking multi-band photometry, allowed us to validate the planetary nature of our candidate (see Sec.~\ref{sec:conclusions}).

\section{Planet Validation}\label{sec:validationmain}
\subsection{TESS data validation}
TOI-1743 was observed by TESS across a total of 39 sectors within 2-minute cadences. A Data Validation (DV) report \citep{2018PASP..130f4502T,2019PASP..131b4506L} covering the longest time span was generated for sectors 14–78, encompassing 32 sectors. Analyzing 158 transits, the measured transit depth was 1974.5$\pm$116.3 ppm with a signal-to-noise ratio (S/N) of 23.3 and a period of 4.26605$\pm$0.00001 days. The comparison of odd and even transit depths was consistent within 0.43$\sigma$. Additional validation tests including the bootstrap test, centroid offset analysis, and ghost test, were all successfully passed.

TOI-5799 was observed during TESS sectors 54 and 81, with 2-minute cadence light curves and DV files were generated for both sectors. The inner planet was reported in the sector 54 DV report with a period of 4.16454$\pm$0.00075 days, a transit depth of 2612.2$\pm$212.5 ppm and a signal-to-noise ratio (S/N) of 13.2. The comparison between odd and even transit depths was consistent within 1.14$\sigma$. Additional validation tests including the bootstrap test, centroid offset analysis, ghost test and difference-imaging centroid tests were successfully passed. 
However, two transits of TOI-5799 c were not reported in the sector 54 DV, possibly due to a low S/N. The first transit of TOI-5799 c was detected in the DV file for sector 81 but because both planets have similar transit depths, the second transit of the outer planet was confused with that of the inner planet. As a result, the orbital period of TOI-5799 c could not be accurately determined.

TOI-6223 was observed during TESS sectors 57 and 84. The Quick Look Pipeline (QLP, \citealt{2020RNAAS...4..204H,2020RNAAS...4..206H}) was applied to full frame images (FFI) and transits of TOI-6223.01 were detected. The SPOC pipeline produced a 2-minute cadence light curve for sector 84 and a corresponding DV report was generated. The reported transit depth was 7782.4$\pm$409.8 ppm with corresponding S/N of 19.2 and an orbital period of 3.85556$\pm$0.00040 days. The odd-even transit depth comparisons showed agreements within 0.15$\sigma$. Additional validation tests including the bootstrap test, centroid offset analysis, ghost test and difference-imaging centroid tests were successfully passed. 

\subsection{Statistical validation}
\label{sec:validation}

We used the tool for Rating Interesting Candidate Exoplanets and Reliability Analysis of Transits Originating from Proximate Stars  ({\tt TRICERATOPS}\footnote{{\tt TRICERATOPS:}~\url{https://github.com/stevengiacalone/triceratops}}, \citealt{Giacalone_2021AJ}) package to compute the false positive probability (FPP), allowing us to assess whether a candidate is a planet or a false positive from a nearby source. {\tt TRICERATOPS} allows us to compute the false positive probability (FPP) and the nearby false positive probability (NFPP).
It uses the phase-folded \emph{TESS} light curves as well as the high-contrast imaging observations in order to improve our results. The transit events were detected on the target stars, which means that NFPP=0 for all candidates.  
We found that FPP$ = (2.996 \pm 1.910)\times10^{-5}$ for TOI-1743.01, FPP$=(1.309 \pm 2.215)\times10^{-5}$ for TOI-6223.01, FPP$= (5.317 \pm 2.983)\times10^{-5}$ for TOI-5799.01, and FPP$= (8.29 \pm 2.74)\times10^{-5}$ for TOI-5799.02. Therefore, all candidates are validated as planets. Moreover, TOI-5799.01 was also statistically validated by \cite{2024PASA...41...30M} using the Sector 54 TESS light curve and a contrast curve obtained from a high-resolution image taken with the Palomar telescope.

\subsection{Archival imaging}
The host stars in this study are relatively nearby and hence have relatively high proper motions (TOI-1743: 200 mas/year, TOI-5799: 301.9 mas/year), except for TOI-6223, which has a proper motion of 32.55 mas/year, allowing us to investigate their current position from archival images to determine whether there are background sources bright enough to produce a transit signal or skew the results obtained from the global analysis.

\begin{figure}[!]
    \centering
    \includegraphics[width=1\linewidth]{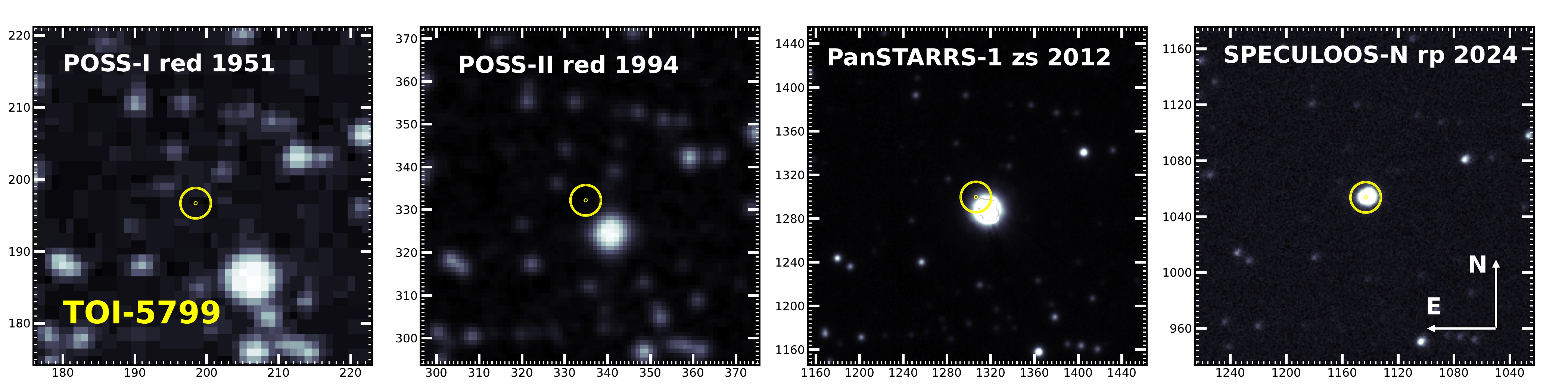}
    \includegraphics[width=1\linewidth]{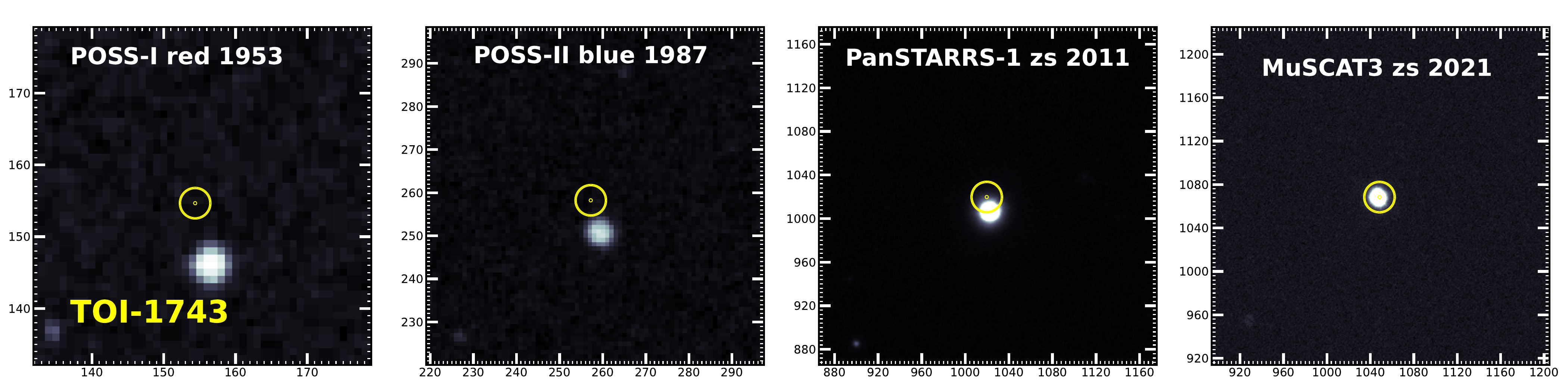}
    \includegraphics[width=1\linewidth]{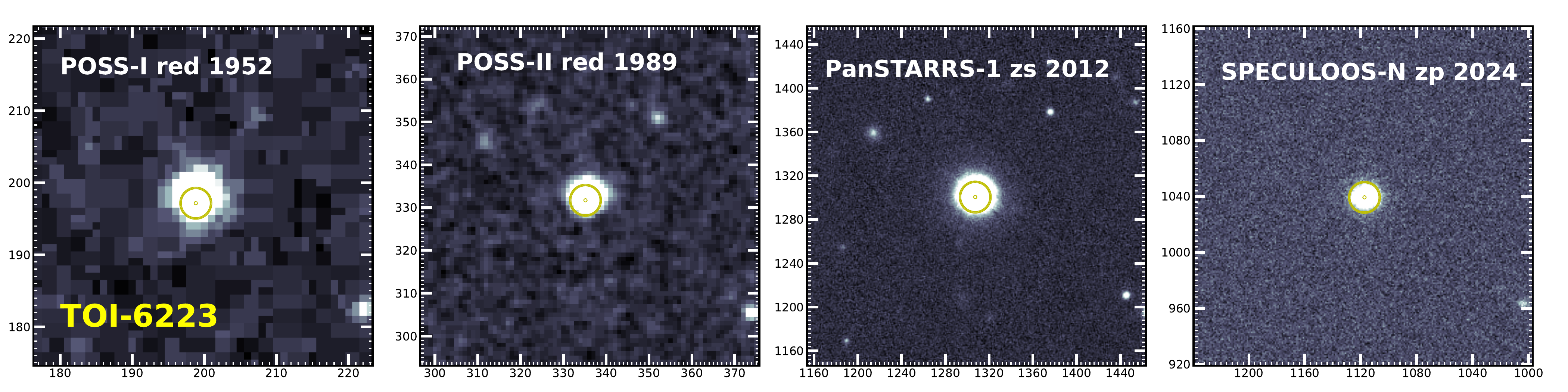}
    \caption{Archival imaging of TOI-5799, TOI-1743, and TOI-6223 (from POSS-I/DSS, POSS-II/DSS2, PanSTARRS1, MuSCAT3 and SPECULOOS-N), shown from top to bottom, respectively. The yellow circle shows the current positions of the targets. 
    }
    \label{fig:Archival_imag}
\end{figure}

We obtained archival images from POSS-I/DSS \citep{1963POSS-I} taken between 1951-1953 in the red filter, POSS-II/DSS \citep{1996DSS_POSS-II} taken between 1987-1994 in the red and blue filters and PanSTARRS-1 \citep{2016arXiv161205560C} taken between 2011-2012 in zs filter to track background stellar sources, as shown in Figure~\ref{fig:Archival_imag}. These images allow us to confirm the absence of any background contaminants along the line of sight within detection limits for TOI-1743 and TOI-5799. For TOI-6223; however, it is not possible to rule out a background star from this diagnostic alone. Nevertheless, such scenario is unlikely as we have ruled out any close companion star within diffraction limits (see Sect. \ref{SAI_images}).

\subsection{Photometric follow-up observations}\label{sec:ground_based_transit}
Ground-based photometric observations were conducted with two primary objectives: to verify the source of the signal and to assess the wavelength dependence of the transit depth\footnote{Note: In the EXOFASTv2 notation, the transit depth is the flux decrement at mid-transit, which depends on the limb darkening (LD). We use it here as (R${_p}$/R$_{\star}$)$^2$ which is independent from LD.}. Nearby eclipsing binaries within the TESS aperture can mimic transit signals (e.g. \citealt{2018AJ....156..234C}), making it essential to confirm if the observed signal originates from the target star itself. Additionally, if the signal is caused by an exoplanet transit, the transit depth should be achromatic across different wavelengths (excluding absorption in the exoplanet atmosphere), as the night-side flux of the exoplanets is generally negligible. The closest neighboring star to TOI-1743 is TIC 219860286 at $44\arcsec$ with $T_{\rm mag}$ of 18.26, and a $\Delta T_{\rm mag}$ of 5.83. The closest neighboring star to TOI-5799 is TIC 328081257 at $19.7\arcsec$ with $T_{\rm mag}$ of 16.8, and a $\Delta T_{\rm mag}$ of 5.63. The closest neighboring star to TOI-6223 is TIC 366038111 at $47.7\arcsec$ with $T_{\rm mag}$ of 17.6, and a $\Delta T_{\rm mag}$ of 4.92.

\begin{figure}[!]
    \centering
    \includegraphics[width=1\linewidth]{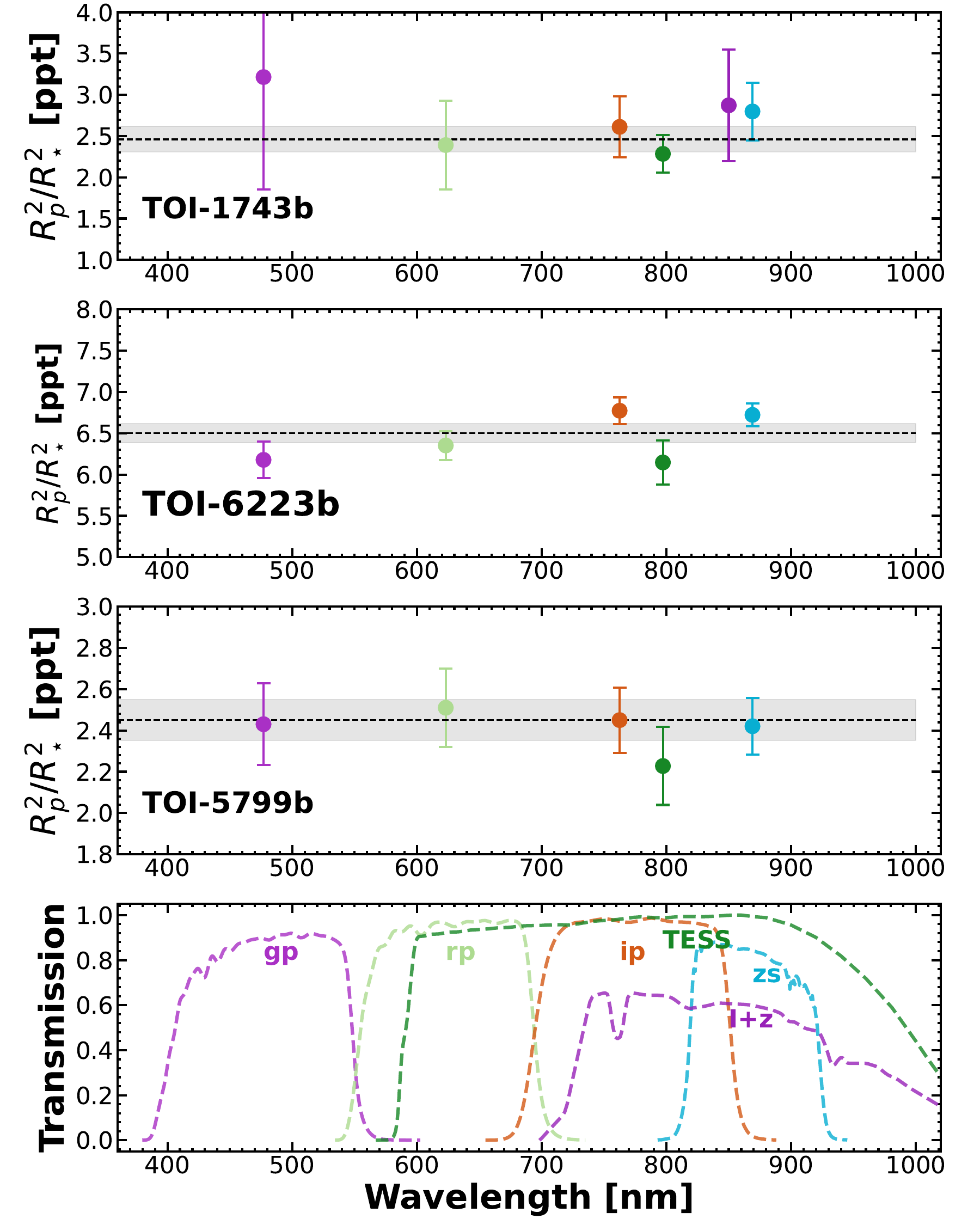}
    \caption{Transit depths independent from limb darkening, $(R_p /R_\star )^2$ measured from different filters (colored points with error bars) for TOI-1743\,b, TOI-6223\,b, and TOI-5799\,b (from top to bottom panels). Global model value using all light curves (i.e. achromatic fit) and its $1\sigma$ uncertainty are shown with dashed line and gray shaded region respectively. Bottom panel shows the transmission curve for related filters.}
    \label{fig:Transit_depth}
\end{figure}

Some of the ground-based photometric observations were performed within seeing limits and aperture sizes of only a few arcseconds were used for photometry to confirm the transit events on the expected target stars. Most of the observations utilized the defocusing technique \citep{2009MNRAS.396.1023S} to achieve better precision but uncontaminated\footnote{Based on stars detected by Gaia which are bright enough to cause measured amplitude by TESS.} apertures used for photometry in these observations. Observations were conducted using multiple filters, covering a wavelength range from 400 to 1100 nm. For each system, we grouped the light curves in the same filters and modeled them simultaneously with the inclination prior from our global model as a GP. The measured transit depths from different filters are in 1~$\sigma$ agreement with the depths from our global models. The measured transit depths are shown in Figure~\ref{fig:Transit_depth}.

\section{Analysis}\label{sec:analysis}
\subsection{SED Fitting}
In order to determine planetary parameters from transits (and radial velocity measurements), absolute parameters of the host stars are required \citep{2010exop.book...55W}. 
Stellar radii ($R_{\rm *}$) can be accurately derived from bolometric fluxes using an empirical approach \citep{2017AJ....153..136S} with the aid of parallax information provided by Gaia \citep{gaia2016, Gaia_Collaboration_2021AandA, gaia2021a}.

To model the bolometric fluxes, we collected broadband apparent magnitudes of the host stars and fitted their spectral energy distributions (SEDs) using the bolometric correction grid from Modules for 
Experiments in Stellar Astrophysics (MESA) Isochrones and Stellar Tracks (MIST) \citep{choi2016}, embedded in {\sc EXOFASTv2} \citep{eastman2019}. We retrieved J, H, Ks magnitudes from 2MASS \citep{cutri2003}, 
W1–W3 magnitudes from AllWISE \citep{cutri2013}, and B, V, g', r', i' magnitudes from APASS9 \citep{henden2016} catalogues for all stars. Additionally, we used the W4 magnitude for TOI-5799 and TOI-1743, gPS, iPS, rPS, yPS, and 
zPS magnitudes from Pan-STARRS \citep{2016arXiv161205560C} for TOI-1743 and TOI-6223.

During the fitting process, the maximum value of interstellar extinction ($A_{\rm V}$) along the line of sight was limited by the value provided by \cite{schlegel1998}. 
We used stellar effective temperature ($T_{\rm eff}$) and metallicity ([Fe/H]) as Gaussian priors (GP) to alleviate $T_{\rm eff}$ - $A_{\rm V}$ degeneracy. For TOI-5799 and TOI-1743, $T_{\rm eff}$ and [Fe/H] values were taken from \cite{2022ApJ...927..122H}; 
for TOI-6223, from \cite{2022ApJS..260...45D}. Finally, we adopted the parallax value from Gaia Data Release 3 (DR3), after adding the offset calculated following \cite{Lindegren2021}. As a result, we obtained the $R_{\rm *}$ values for all host stars to be used in the global modeling as GP. $R_{\rm *}$ and other parameters derived from SED fitting ($A_{\rm V}$ and $T_{\rm eff}$) are listed in Table \ref{stellarpar} and SED models are shown in Figure~\ref{fig:SED}.

\begin{table*}[!]
\caption{Stellar properties of TOI-1743, TOI-5799, and TOI-6223 from astrometry, photometry, and spectroscopy.}
\centering
	{\renewcommand{\arraystretch}{1.}
        \resizebox{0.9\textwidth}{!}{
		\begin{tabular}{lccccc}
			\hline
			\hline
			\multicolumn{5}{c}{  Star information}   \\
			\hline
			{\it Target designations:}  & & \\
			            & TOI 1743 & TOI 5799  & TOI 6223  \\
                        & TIC 219860288 & TIC 328081248 & TIC 288144647  \\
			              & GAIA DR3 1651807592698185728 & GAIA DR3 1807887150930343424  & GAIA DR3 2872039996566868480 \\
			            & 2MASS J17105910+7152170  & 2MASS J20063105+1559171 & 2MASS J23434812+3304025  \\
   			Parameter &  &  & &  Source   \\
			\hline
            \hline
			{\it Parallax  and distance:} &   \\
			RA [J2000]     &  17:10:59.36  & 20:06:31.24  &  23:43:48.14 &  (1) \\
			Dec [J2000]    &  +71:52:19.93   & +15:59:20.91 &  +33:04:02.05 &  (1)\\
			Plx [$mas$] & $24.207 \pm 0.017$ & $35.896\pm0.017$ & $7.45\pm0.02$ &    (1)\\
            $\mu_{RA}$ [mas yr$^{-1}$] & $66.75\pm0.02$ & $177.668\pm0.017$ &  $14.015\pm0.024$ & (1) \\
            $\mu_{Dec}$ [mas yr$^{-1}$] & $188.68\pm0.02$ & $244.112\pm0.015$ &  $-29.379\pm 0.12$ & (1) \\
			Distance [pc]  & $41.31\pm0.03$ & $27.858\pm0.013$ &  $134.17\pm0.36$ & (1)\\
			\hline
			\multicolumn{4}{l}{{\it Photometric properties:}} & \\
			TESS$_{\rm mag}$            & $12.425 \pm 0.007$  & $ 11.1798\pm0.0074 $ &  $12.6758 \pm 0.00729$ & (2)  \\
			$V_{\rm mag}$ [UCAC4]       & $14.972 \pm 0.073$  & $ 13.29\pm0.08 $ &  $14.22 \pm 0.103$ & (3) \\
            $B_{\rm mag}$ [UCAC4]       & $16.679 \pm 0.125$  & $14.855 \pm 0.016$ & $15.753 \pm 0.052$ & (3) \\
            $R_{\rm mag}$ [UCAC4]       & $ 13.990 $  & $ 13.530 $ &  $ 13.42 $ & (3) \\
			$J_{\rm mag}$ [2MASS]       & $10.807 \pm 0.021$  & $ 9.742\pm 0.023$ &  $11.504 \pm 0.022$ &  (4) \\
			$H_{\rm mag}$ [2MASS]       & $10.246 \pm 0.017$  & $ 9.192 \pm 0.030  $ &  $10.799 \pm 0.028$ &    (4) \\
			$K_{\rm mag}$ [2MASS]       & $9.999 \pm 0.014$  & $ 8.952 \pm 0.021 $ &  $10.689 \pm 0.021$ &   (4) \\			
			$G_{\rm mag}$ [Gaia DR3]    & $13.729 \pm 0.001$  & $12.3212 \pm 0.0003$  & $13.6201 \pm 0.00028$ &  (1)  \\
			$W1_{\rm mag}$ [WISE]       & $9.840 \pm 0.023$  & $8.806 \pm 0.023$ &  $10.560 \pm 0.023$ &  (5) \\
			$W2_{\rm mag}$ [WISE]       & $9.658 \pm 0.020$  & $8.654 \pm 0.021$ &  $10.593 \pm 0.020$ &   (5) \\
			$W3_{\rm mag}$ [WISE]       & $9.517 \pm 0.027$  & $ 8.591\pm0.023 $  & $10.44 \pm 0.088$ &   (5)\\
            $W4_{\rm mag}$ [WISE]       & $9.584 \pm 0.428$  & $ 8.394\pm 0.256$  & $ 8.58 $ &   (5)\\
			\hline
   			\multicolumn{4}{l}{\it Spectroscopic and derived parameters}  \\
			$T_{\rm eff}$ [K]              &  $  3277^{+78}_{-77}  $ & $ 3452 \pm 96 $ &  $ 3895 \pm 44 $ &   this work\\
			$\log g_\star$ [cgs]           &  $  4.909^{+0.032}_{-0.033} $  & $ 4.936^{+0.034}_{-0.035} $ &  $ 4.682^{+0.025}_{-0.024} $ &  this work\\
			$[Fe/H]$ [dex]                 & $  0.22^{+0.14}_{-0.15}  $  & $ 0.01^{+0.16}_{-0.17} $ & $ -0.12^{+0.14}_{-0.15} $ &  this work\\
			$M_\star$  [$M_\odot$]         & $  0.339^{+0.027}_{-0.031}  $  & $ 0.325 \pm 0.03 $ &  $ 0.582 \pm 0.027 $ & this work\\
			$R_\star$  [$R_\odot$]         & $  0.338 \pm 0.014  $  & $ 0.321 \pm 0.014 $ &  $0.576 \pm 0.022$ & this work\\
			$Av$ [mag]    &  $ 0.021^{+0.038}_{-0.016}$ & $ 0.055^{+0.098}_{-0.041}$ & $ 0.054^{+0.054}_{-0.038}$ & this work\\
			$\rho_\star$  [$g/cm^{3}$]   & $ 12.3^{+1.3}_{-1.2} $  & $ 13.8^{+1.6}_{-1.4} $   & $ 4.29^{+0.42}_{-0.36} $ & this work \\
			$Age$  [Gyr]                   & $9.7^{+2.8}_{-4}$ & $9.6^{+2.8}_{-4.1}$  & $9.8^{+2.7}_{-4.0}$& this work \\
			Spectral type                  & M4V & M2V & M0V&  this work [KAST]\\
                                            & -- & M2V & -- &  this work [MagE]\\
                                            & -- & M2.5$\pm$0.5 & M1.5$\pm$0.5 &  this work [SpeX]\\
            \hline
	\end{tabular} }}
	\tablefoot{ 
	{\bf (1):} Gaia DR3 \cite{Gaia_Collaboration_2021AandA}; 
	{\bf (2)} \emph{TESS} Input Catalog \cite{Stassun_2018AJ_TESS_Catalog}; 
	{\bf (3)} UCAC4 \cite{Zacharias_2012yCat.1322};
	{\bf (4)} 2MASS \cite{Skrutskie_2006AJ_2MASS};
	{\bf (5)} WISE \cite{Cutri_2014yCat.2328}.
	\label{stellarpar}}
\end{table*}

\subsection{Global Modelling}
We simultaneously modeled the ground based transit observations (listed in Table-\ref{obs_table}) along with the TESS observations to better constrain the transit parameters. {\sc EXOFASTv2} can fit transit and stellar isochrones at the same time using MIST as they feed each other. Stellar density ($\rho_{\star}$) derived from transit models (e.g. \citealt{2003ApJ...585.1038S}) can be used in MIST to better constrain the absolute parameters of the host star while stellar parameters can be used to calculate limb darkening parameters following \cite{claret2011}. The effective temperatures of TOI-1743 and TOI-5799 fall below the range given by \cite{claret2011} so we set the limb darkening parameters as free but their bounds are still controlled by the criteria given by \cite{2013MNRAS.435.2152K}.

Since the resulting $T_{\rm eff}$ values from our SED models are in excellent agreement with their priors from previous studies, we used $T_{\rm eff}$ and [Fe/H] values as GP, same as our SED fitting. This time we also used $R_{\rm *}$ as GP from our SED models but its uncertainty was calculated following \cite{tayar2020}. Orbital Period ($P_{\rm orb}$), time of first transit ($T_{\rm 0}$), orbital inclination (\textit{i}) were also supplied as uniform priors between $\pm$$\infty$ to speed up the convergence time.

In {\sc EXOFASTv2}, convergence is controlled by two metrics: The Gelman-Rubbin statistic ($R_{\rm Z}$) which measures the degree of similarity between independent chains and the other metric is the number of independent draws ($T_Z$), defined as the ratio of the chain lengths to their correlation length. We used the default {\sc EXOFASTv2} values for these metrics ($R_{\rm Z}$ < 1.01 and $T_Z$ > 1000) which are the strict recommendations of \cite{2006ApJ...642..505F}. After convergence, resulting median values and their 1-$\sigma$ uncertainties from posterior distributions are listed in Table~\ref{mcmc_params_3}. Ground-based transit observations and global transit models are shown in Figures~\ref{fig:ground_based_obs_1743_5799b}, \ref{fig:ground_based_obs_5938_6223} and \ref{fig:ground_based_obs5799c}.

\begin{table*}[!]
\caption{Median values and 68\% confidence intervals for TOI-1743\,b, TOI-6223\,b,  TOI-5799\,b, and TOI-5799\,c and FPP values.}
	\begin{center}
		{\renewcommand{\arraystretch}{1.2}
				\resizebox{0.99\textwidth}{!}{
			\begin{tabular}{llccccc}
				\hline
			        Parameter & Unit &  TOI-1743\,b  &  TOI-6223\,b &  TOI-5799\,b  & TOI-5799\,c    \\
           \hline
            Orbital period, $P$ & days& $4.266046\pm0.000002$ &  $3.855219 \pm 0.000004$ & $4.164507^{+0.000005}_{-0.000006}$ & $14.010449^{+0.000056}_{-0.000046}$ \\ 
            Transit-timing,  & BJD$_{\rm TDB}$ &  $9575.17230\pm0.00027$  &  $9856.48875\pm0.00077$ & $10255.41850^{+0.00032}_{-0.00030}$&$10383.0352\pm0.0012$  \\
            $T_0 - 2450000$ &&&&& \\
            Orbital semi-major axis, $a$ & au      & $0.03589^{+0.00092}_{-0.00110}$ &  $0.04018^{+0.00061}_{-0.00062}$ & $0.0348^{+0.0010}_{-0.0011}$&$0.0782^{+0.0023}_{-0.0025}$ \\
            Impact parameter, $b$ & --      & $0.827^{+0.016}_{-0.017}$  &    $0.345^{+0.066}_{-0.096}$ & $0.346^{+0.088}_{-0.14}$&$0.624^{+0.052}_{-0.061}$ \\
			Transit duration, $T_{14}$ & days            &  $0.03848^{+0.00080}_{-0.00075}$  &  $0.08392^{+0.00061}_{-0.00060}$ & $0.05634^{+0.00081}_{-0.00077}$&$0.0721^{+0.0026}_{-0.0027}$  \\
			Orbital inclination, $i$ & deg         &  $87.92\pm0.11$ &   $88.68^{+0.39}_{-0.30}$ & $89.15^{+0.35}_{-0.25}$&$89.317^{+0.082}_{-0.077}$ \\
			Radius ratio, $R_p /R_\star $&--& $0.0496^{+0.0016}_{-0.0015}$ &  $0.08065\pm0.00071$ & $0.04951^{+0.00092}_{-0.00094}$&$0.0501\pm0.0017$\\
            Scaled semi-major axis, $a/R_\star$          &--&   $22.81^{+0.78}_{-0.74}$   &    $14.99^{+0.47}_{-0.43}$ & $23.31^{+0.85}_{-0.82}$&$52.3^{+1.9}_{-1.8}$  \\
            Planet radius, $R_p$ & $R_\oplus $    &  $1.83^{+0.11}_{-0.10}$ &    $5.07^{+0.22}_{-0.22}$ & $1.733^{+0.096}_{-0.090}$&$1.76^{+0.11}_{-0.10}$\\
            False Positive Probability, FPP &--&  $(2.996 \pm 1.910)\times10^{-5}$ &    $(1.309 \pm 2.215)\times10^{-5}$ & $(5.317 \pm 2.983)\times10^{-5}$&$(8.29 \pm 2.74)\times10^{-5}$\\
				\hline
			        Estimated Parameters &  &    &  \\
           \hline
            $^a$Equilibrium temperature, $T_{\rm eq}$ & K    &  $485^{+14}_{-13}$ &   $714\pm14$ &  $505\pm16$&$337\pm11$  \\
            $^b$RV semi-amplitude, $K_{\star}$ & $m/s $    &  $3.7^{+1.4}_{-0.9}$ &   $14.9^{+5.3}_{-3.4}$ & $3.5^{+1.3}_{-0.8}$&$2.4^{+0.9}_{-0.6}$\\
            $^b$Planet mass, $M_p$ & $M_\oplus $    &  $4.6^{+1.7}_{-1.1}$ &  $25.6^{+9.2}_{-6.0}$ & $4.1^{+1.5}_{-1.0}$&$4.2^{+1.6}_{-1.0}$\\
            $^c$TSM & -- & $60\pm25$  &$62\pm18$ & $87\pm36$ & $60\pm26$\\
   \hline
		\end{tabular}}}
	\end{center}
	\tablefoot{$^{(a)}$Assuming no albedo and perfect redistribution.$^{(b)}$ Uses measured radius and estimated mass from \citet{Chen2017} $^{(c)}$ TSM values are computed from \cite{2018PASP..130k4401K}.}
	\label{mcmc_params_3}
\end{table*}

\section{Discussion}\label{sec:concanddiscs}
\subsection{On The Planet Validation}
\label{sec:conclusions}
Using archival images, diffraction limited high resolution imaging, and seeing limited ground-based transit observations, we confirm that the transit signals originate from the target stars. For TOI-6223, it is not possible to detect foreground or background objects using archival images due to the star's low proper motion. However, the SED of TOI-6223 is consistent with a single star. Moreover, the renormalized unit weight error (RUWE) value from Gaia supports the single-star hypothesis (e.g. \citealt{2021ApJ...907L..33S}). 

Using ground-based observations, we show that the signals are achromatic across the observed wavelengths for all targets except TOI-5799\,c, for which we lack sufficient data. However, the $a/R_\star$ values obtained from individual transit models of the TOI-5799 planets yield consistent stellar density ($\rho_\star$) estimates (see \citealt{2003ApJ...585.1038S} for $a/R_\star$ - $\rho_\star$ relation), supporting the interpretation that the signals originate from transits of two distinct planets in the same system. Finally, our statistical analysis using {\tt TRICERATOPS} indicates that the false positive probabilities (FPP) of the signals are practically zero. The measured radii of all candidates fall below the minimum brown dwarf radius (e.g. \citealt{2013ApJ...767...77S}), and TRES radial velocity measurements confirm the absence of a close stellar companion. Therefore, we confirm that the signals detected by TESS are consistent with exoplanet transits.

\subsection{Characterization Prospects}
\label{section:cp}
Increasing the sample size of fully characterized exoplanets is essential for improving our understanding of their formation history and evolutionary processes. Precise measurements of planetary mass, bulk density and eccentricity provide valuable insights into where and how these planets formed. These key parameters can be derived from combination of high resolution radial velocity and transit observations (e.g. \citealt{2010exop.book...15M}). Accurately determining these parameters demands high-precision spectroscopic observations, which can only be achieved using stabilized spectrographs like MAROON-X \citep{2018SPIE10702E..6DS} mounted on the 8.1m Gemini-North telescope. For exposure times of 2700 seconds and good observing conditions, MAROON-X exposure time calculator predicts an radial velocity (RV) precision of 0.7, 0.5 and 0.9 m/s for TOI-1743, TOI-5799 and TOI-6223 respectively. For these systems, we calculated radial velocity semi-amplitudes using mass-radius relation from \cite{Chen2017} as listed in Table~\ref{mcmc_params_3}, which are many sigma higher than precisions calculated for single observations. Furthermore, we analyzed the TESS light curves to search for stellar rotation via spot modulation using the methods of \cite{2023AJ....166...16P, 2022ApJ...936..109P}, but found no statistically significant rotational modulation signals for TOI-5799 or TOI-6223. The non-detection of rotation signal may indicate that these stars are inactive, which is consistent with our findings from spectral analysis (see Sec. \ref{sec:Shane}, \ref{sec:Magellan} and \ref{sec:IRTF}) and facilitates to accurately measure RVs. For TOI-1743, \cite{2020ApJS..250...20C} searched for the stellar rotation period using five sectors of TESS light curves and reported no detection due to the high noise level. We analyzed 36 sectors of TESS data and found a tentative signal with a period of 4.56 days with an amplitude of 0.255 ppt. Although the amplitude is very low, a 4.56 day period indicates that the star is still active and it is consistent with star's modest H$_\alpha$ emission (See Sec. \ref{sec:Shane}). If the signal originates from the target star, radial velocity follow-up would be complicated because the stellar rotation period is close to the planetary orbital period, making it difficult to separate stellar activity from the planetary signal. Alternatively, the signal may arise from contamination by nearby stars, which could produce the observed variability even with minor contamination given its small amplitude.

\begin{figure}
    \centering
    \includegraphics[width=1\columnwidth]{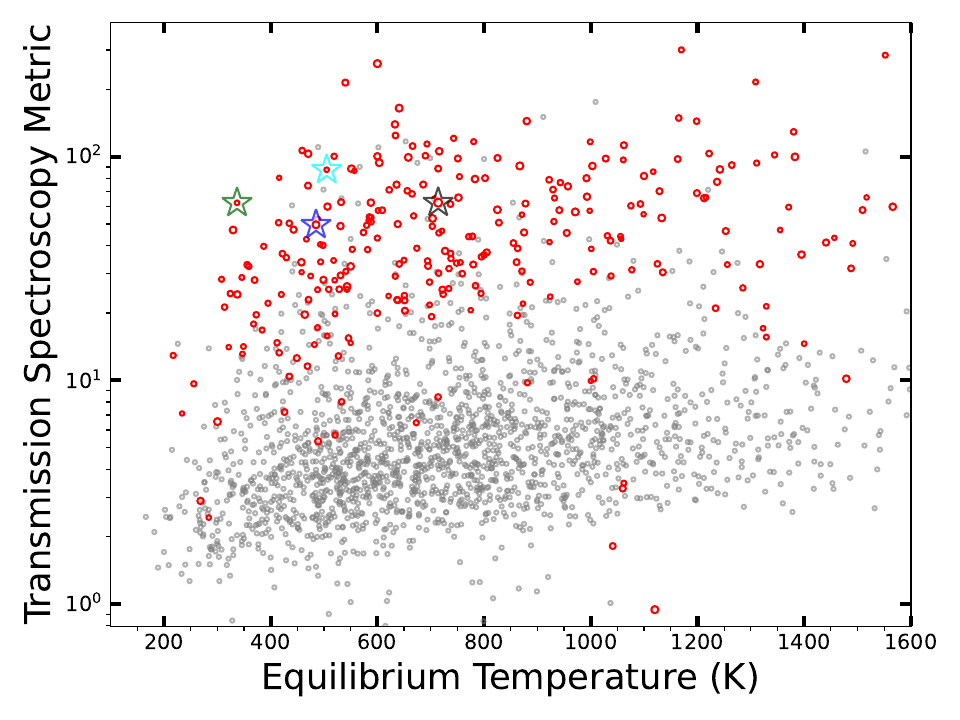}
    \caption{
        Transmission spectroscopy metric as a function of equilibrium temperature of confirmed transiting sub-Neptune exoplanets (gray circles). Confirmed TESS planets are shown with red circles. Marker sizes for TESS planets are proportional to the exoplanet radii. TOI-5799 c, TOI-1743 b, TOI-5799 b and TOI-6223 b denoted with green, blue, cyan and black stars respectively. Data were extracted from  {\tt NASA Exoplanet Archive.}    
}
    \label{fig:TSM}
\end{figure}

Although the mass and radius are known, the bulk density of exoplanets can match a variety of interior compositions \citep{2019PNAS..116.9723Z}. Exploration of their atmospheric compositions may alleviate the degeneracies between interior composition models \citep{2022ASSL..466....3R}. It is possible to probe the atmosphere of such small exoplanets with the James Webb Space Telescope (JWST) using transmission spectroscopy technique (e.g. \citealt{2025ApJ...983L..40M}). We calculated the transmission spectroscopy metric (TSM, \citealt{2018PASP..130k4401K}) which is proportional to the expected transmission spectroscopy signal-to-noise ratio (SNR) for JWST observations as given in Table~\ref{mcmc_params_3}. All exoplanets within this study have relatively high TSM values among sub-Neptune planets as shown in Figure~\ref{fig:TSM}. Especially, TOI-5799\,c has the highest TSM among temperate ($T_{\rm eq}$ < 400 K) sub-Neptune planets in orbit around low mass stars ($M_\star$ $\leq 0.65$ $M_\odot$, $T_{\rm eff}<$  4700 K). This could also make TOI-5799\,c one of the most favorable targets among HZ exoplanets with Venus-like irradiation. This planet has thus a high astrobiological importance. However, in order to estimate whether the planet is in the HZ, planetary mass determination is required \citep{2013ApJ...767L...8K}. 

To further investigate the feasibility of characterizing the atmospheres, we adopted the same procedure as \cite{2024A&A...687A.264B}. First, we simulated JWST observations for each target using the online tool \texttt{PandExo} \citep{2017PASP..129f4501B}. JWST instruments, exposure times and predicted SNR values for the simulation are given in Table \ref{tab:ExpSNR}. Then we modeled the simulated observations using PLanetary Atmospheric Transmission for Observer Noobs \citep[\texttt{PLATON};][]{2019PASP..131c4501Z,2020ApJ...899...27Z}. We identified several $\rm H_2O$ lines in the near infrared (between 0.8-2.8 $\mu$m), $\rm CH_4$ lines at 0.8, 1.0, 1.3, 1.7, 2.3, and 3.3 $\mu$m and $\rm CO_2$ signatures around 4.5 $\mu$m in synthetic transmission spectra of all four exoplanets. We also detected potential traces of $\rm H_2S$ lines at 1.9 and 3.7 $\mu$m and $\rm CO$ at 4.50 $\mu$m for TOI-6223\,b. Synthetic spectra and best fit models are shown in Figure~\ref{fig:STS}.

We simulated the TOI-5799 system to quantify the expected amplitudes of transit timing variations (TTVs), with the aim of assessing whether these methods could provide useful dynamical constraints for the system. We used the \texttt{REBOUND} N-body integrator \citep{Rebound2012A&A...537A.128R}, employing its \texttt{IAS15} adaptive, high-precision integrator \citep{IAS152015MNRAS.446.1424R}, to simulate the system for a 450-day timespan with a time resolution of approximately 13 minutes. From the simulated orbital trajectories, we determined the exact times of transit using a Newton-Raphson root-finding method, implemented with the \texttt{scipy} package \citep{SCIPY2020SciPy-NMeth}. Specifically, we computed the four contact points as well as the mid-transit times for each event. For planet b, we identified 108 transits, with a measured full TTV amplitude of 4.46 seconds. For planet c, we calculated 32 transits, with a full TTV amplitude of 5.97 seconds. These results suggest that the expected TTV signals in the TOI-5799 system are too small to be detectable with the available photometric data.

\subsection{Populating Rare Spots in P--\texorpdfstring{$R_p$}{Rp} Plane}
\subsubsection{Radius Valley for sub-Neptunes}
\begin{figure}
    \centering
    \includegraphics[width=\columnwidth]{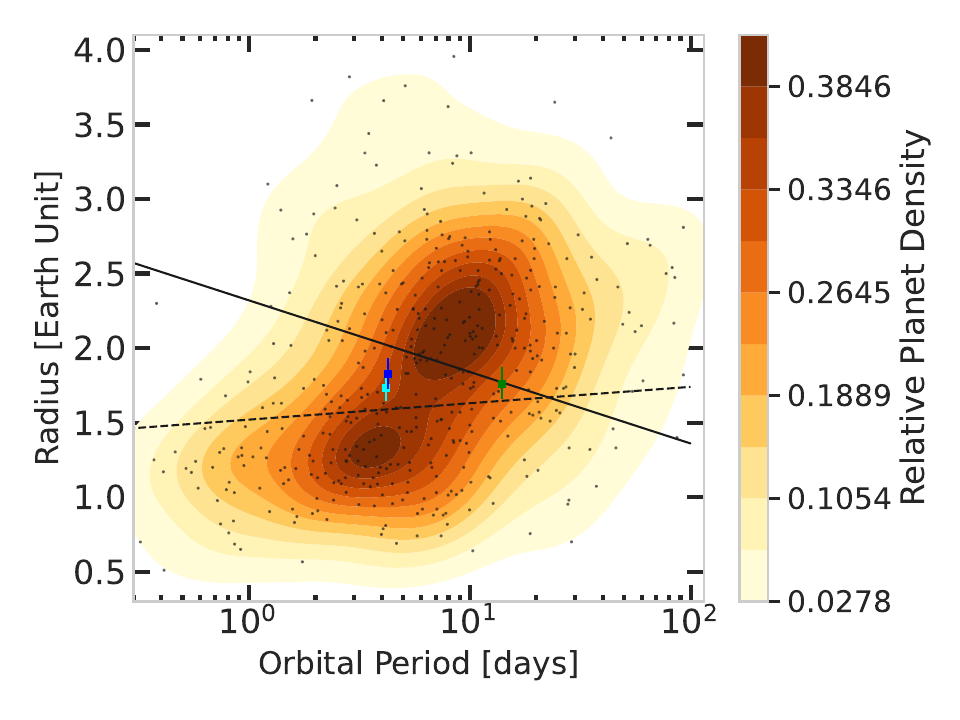}
    \caption{
        Radii of sub-Neptune exoplanets with P $\leq$ 100 days, around low mass stars ($T_{\rm eff}<4700 K$) as a function of orbital period. TOI-1743\,b, TOI-5799\,b and TOI-5799\,c shown with blue, cyan and green squares respectively. Dashed black line shows the radius valley of sub-Neptune exoplanets around low mass stars measured by \cite{2020AJ....159..211C} while black line shows the radius valley for exoplanets around Sun-like stars measured by \cite{2019ApJ...875...29M}, scaled to low mass stars by \cite{2020AJ....159..211C}. Only exoplanets with precisely measured radii (>10$\sigma$) were included. Data were extracted from  {\tt NASA Exoplanet Archive.}
}
    \label{fig:rad_valley}
\end{figure}

The bimodal radius distribution of sub-Neptune planets around Sun-like stars is also evident among low mass stars ($M_\star$ < 0.65 $M_\odot$) with the first peak at $\sim$0.9–1.4 $R_\oplus$ and the second at $\sim$1.9–2.3 $R_\oplus$ \citealt{2020AJ....159..211C}). However, the slope of the radius valley in period-radius (P-$R_p$) and insolation flux-radius (F-$R_p$) plane has opposing signs for Sun-like stars and low mass stars \citep{2020AJ....159..211C, 2019ApJ...875...29M, 2018MNRAS.479.4786V}. The slopes for low mass and Sun-like stars (scaled for low mass stars, see \citealt{2020AJ....159..211C}) intersect at P = 23.52 days. The region between these slopes has been identified as a key area of interest for investigating the rocky-to-non-rocky transition around low-mass stars \citep{2020AJ....159..211C}. With the radii of $1.83^{+0.11}_{-0.10}$ $R_\oplus$, $1.733^{+0.096}_{-0.090}$ $R_\oplus$ and $1.76^{+0.11}_{-0.10}$ $R_\oplus$, TOI-1743\,b, TOI-5799\,b and TOI-5799\,c fall in the P$\leq$23.52 day-side of this region of interest as shown in Figure~\ref{fig:rad_valley}. \cite{2020AJ....159..211C} argues that such planets could be rocky and subject to the gas-poor formation, or they may have a dense core and be subject to core-powered mass loss. These scenarios can be robustly tested through bulk density measurements derived from high-precision radial velocity observations.

\subsubsection{Neptune Desert}
The distribution of close-in planets reveals an interesting feature: the dearth of short-period Neptunian exoplanets, commonly referred to as the Neptunian desert or sub-Jovian desert \citep{2011A&A...528A...2B, 2011ApJ...727L..44S, 2011ApJ...742...38Y,2013ApJ...763...12B,2016NatCo...711201L,2016A&A...589A..75M}. More recently, \cite{2024A&A...689A.250C} identified three regions in the P-$R_p$ plane: (1) the Neptunian desert, corresponding to planets with orbital periods P$\leq$3.2 days, where Neptunian planets are notably scarce; (2) the Neptunian savanna, a sparsely populated region for exoplanets with P$\geq$5.7 days and (3) the Neptunian ridge, located between these two regions, where there is a sharp increase in the population of Neptunian planets.

\begin{figure}[ht!]
    \centering
    \includegraphics[width=1\columnwidth]{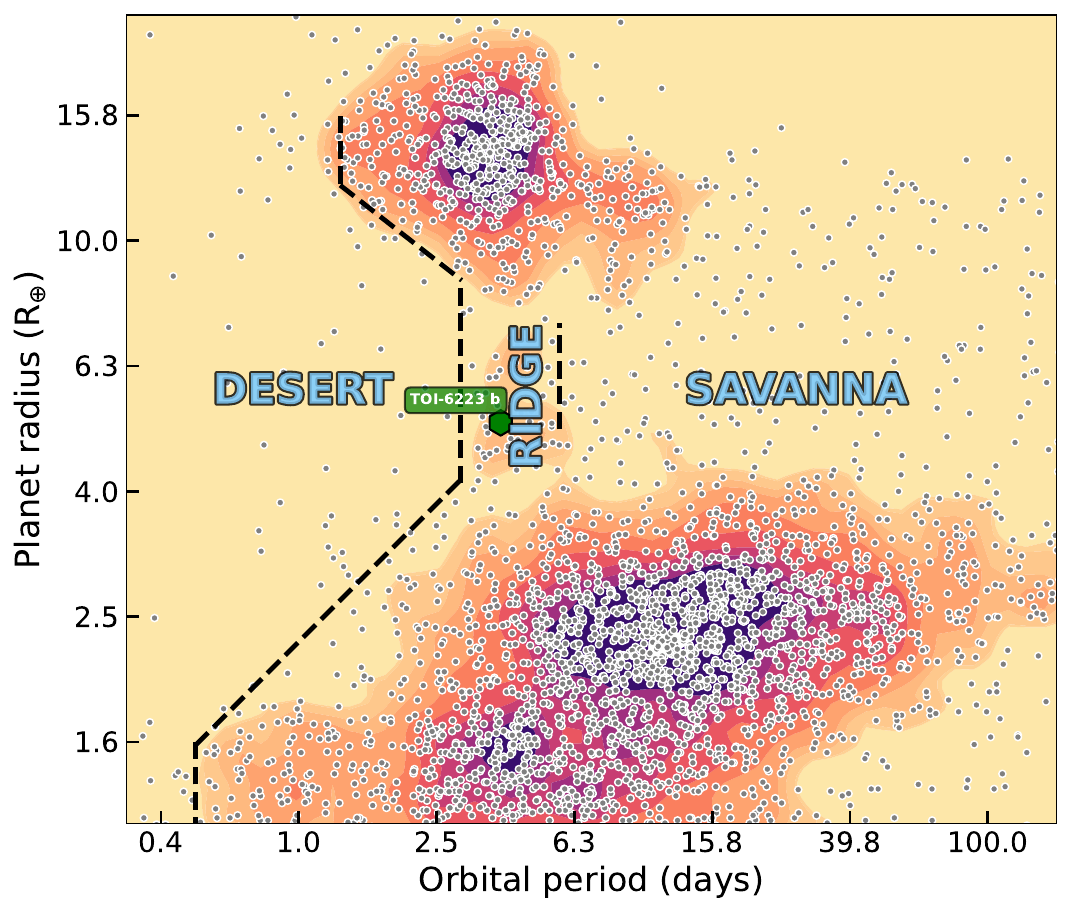}
    \caption{
        Period-radius diagram of transiting exoplanets. The boundaries of the Neptunian desert, ridge, and savanna \citep{2024A&A...689A.250C} are indicated with black dashed lines. TOI-6223\,b is shown with green hexagon. Data were extracted from  {\tt NASA Exoplanet Archive.} This plot was generated with {\tt nep-des}
(https://github.com/castro-gzlz/nep-des).
}
    \label{fig:nep_des}
\end{figure}

With a radius of $5.12^{+0.24}_{-0.25}$ $R_\oplus$ and an orbital period of 3.86 days, TOI-6223\,b populates the Neptunian ridge as shown in Figure~\ref{fig:nep_des}. \cite{2024A&A...689A.250C} suggests that Neptunian planets on the ridge reached their current positions through HEM and are expected to retain non-zero eccentricities. This prediction can be tested for TOI-6223\,b as its host star is suitable for precise radial velocity measurements as discussed in Section \ref{section:cp}.

\section{Conclusions}\label{sec:concs}
We have presented the discovery and validation of four transiting exoplanets: TOI-1743\,b, TOI-5799\,b, TOI-5799\,c, and TOI-6223\,b.
TOI-1743\,b, TOI-5799\,b and TOI-6223\,b were detected by TESS and TOI-5799\,c was detected by \texttt{SHERLOCK} and validated through extensive ground-based follow-up.
TOI-1743\,b, TOI-5799\,b, and TOI-5799\,c are super-Earths with radii near 1.7--1.8\,$R_\oplus$, while TOI-6223\,b is a slightly larger Neptune-sized planet with a radius of ${\sim}5.1\,R_\oplus$.
These planets orbit bright early M dwarfs ($K < 11$), making them ideal candidates for precise characterization. In fact, TOI-5799\,b is already included in a characterization survey:  Origins, Compositions, and Atmospheres of Sub-Neptunes (OrCAS, \citealt{2025AJ....169...89C}).

These new systems add valuable targets to the population of small planets in key regions of the period--radius diagram.
Notably, TOI-1743\,b, TOI-5799\,b, and TOI-5799\,c lie within the radius valley, providing insight into the processes that sculpt the bimodal distribution of sub-Neptune exoplanets, including photoevaporation and core-powered mass loss.
Meanwhile, TOI-6223\,b lies on the Neptunian ridge, a transition region that may record migration pathways that are distinct from those that populate the Neptunian desert and savanna.

The brightness and proximity of the host stars make these systems excellent candidates for future radial velocity monitoring to measure precise masses and eccentricities as well as for atmospheric characterization efforts with JWST.
For radial velocity follow-up, the expected semi-amplitudes range from ${\sim}$2--4\,m\,s$^{-1}$ for the super-Earths to ${\sim}$15\,m\,s$^{-1}$ for the Neptune-sized TOI-6223\,b, well within reach of current NIR spectrographs.
Using \texttt{PandExo} and \texttt{PLATON}, we simulated transmission spectra for each planet to assess the feasibility of characterizing their atmospheres with JWST.
We found that the prominent molecular signatures of H$_2$O, CH$_4$, and CO$_2$, and potentially H$_2$ and CO, should be detectable in the near- and mid-infrared at high SNR.
Given this accessibility, TOI-5799\,c in particular, with an orbital period of ${\sim}14$\,d and an equilibrium temperature near the inner edge of its star's habitable zone, represents an intriguing target for studying temperate super-Earths.

Together, these discoveries expand the small but growing sample of well-characterized planets in key regions of the period--radius diagram.
Continued follow-up---including precise mass measurements and JWST transmission spectroscopy---will help clarify the origins of the radius valley and the structure of the Neptunian ridge and desert.
Such efforts will help to deepen our understanding of how close-in planets form, evolve, and retain their atmospheres (or not) around low-mass stars.

\bibliographystyle{aa}
\bibliography{aa.bib}

\appendix
\section{Acknowledgements}
\small{SY acknowledges support from the T\"UBİTAK 2214-A program.
In this study, the observational data obtained within the scope of project numbered 22BT100-1958 and 25ATUG100-3012 carried out using the TUG100 telescope at the TUG (T\"UBİTAK National Observatory, Antalya) site of the T\"urkiye National Observatories has been utilised, and we express our gratitude for the invaluable support provided by the T\"urkiye National Observatories, the observation team, and all its staff.
The data in this study were obtained with the T80 telescope at the Ankara University Astronomy and Space Sciences Research and Application Center (Kreiken Observatory) with the project number of 24B.T80.07.
Funding for KB was provided by the European Union (ERC AdG SUBSTELLAR, GA 101054354).
I.A.S. acknowledges the support of M.V. Lomonosov Moscow State University Program of Development.
This material is based upon work supported by the National Aeronautics and Space Administration under Agreement No.\ 80NSSC21K0593 for the program ``Alien Earths''.
The results reported herein benefited from collaborations and/or information exchange within NASA’s Nexus for Exoplanet System Science (NExSS) research coordination network sponsored by NASA’s Science Mission Directorate.
We acknowledge financial support from the Agencia Estatal de Investigaci\'on of the Ministerio de Ciencia e Innovaci\'on MCIN/AEI/10.13039/501100011033 and the ERDF “A way of making Europe” through project PID2021-125627OB-C32, and from the Centre of Excellence “Severo Ochoa” award to the Instituto de Astrofisica de Canarias. This work is partly supported by JSPS KAKENHI Grant Numbers JP24K17082, JP24K00689, JP24H00248, JP24H00017, JP21K13955, JST SPRING, Grant Numbers JPMJSP2108, JSPS Grant-in-Aid for JSPS Fellows Grant Number JP25KJ1036, and JSPS Bilateral Program Number JPJSBP120249910. This article is based on observations made with the MuSCAT2 instrument, developed by ABC, at Telescopio Carlos Sánchez operated on the island of Tenerife by the IAC in the Spanish Observatorio del Teide. This paper is based on observations made with the MuSCAT3 instrument, developed by the Astrobiology Center and under financial supports by JSPS KAKENHI (JP18H05439) and JST PRESTO (JPMJPR1775), at Faulkes Telescope North on Maui, HI, operated by the Las Cumbres Observatory.
DRC acknowledges partial support from NASA Grant 18-2XRP18\_2-0007. This research has made use of the Exoplanet Follow-up Observation Program (ExoFOP; DOI: 10.26134/ExoFOP5) website, which is operated by the California Institute of Technology, under contract with the National Aeronautics and Space Administration under the Exoplanet Exploration Program. Based on observations obtained at the Hale Telescope, Palomar Observatory, as part of a collaborative agreement between the Caltech Optical Observatories and the Jet Propulsion Laboratory operated by Caltech for NASA. The Observatory was made possible by the generous financial support of the W. M. Keck Foundation. The authors wish to recognize and acknowledge the very significant cultural role and reverence that the summit of Maunakea has always had within the Native Hawaiian community. We are most fortunate to have the opportunity to conduct observations from this mountain.
This paper includes data gathered with the 6.5 meter Magellan Telescopes located at Las Campanas Observatory, Chile.
Visiting Astronomer at the Infrared Telescope Facility, which is operated by the University of Hawaii under contract 80HQTR24DA010 with the National Aeronautics and Space Administration.
This work made use of \texttt{TESS-cont} (\url{https://github.com/castro-gzlz/TESS-cont}), which also made use of \texttt{tpfplotter} \citep{2020A&A...635A.128A} and \texttt{TESS-PRF} \citep{2022ascl.soft07008B}.
Funding for the TESS mission is provided by NASA's Science Mission Directorate. KAC and CNW acknowledge support from the TESS mission via subaward s3449 from MIT.
This paper made use of data collected by the TESS mission, obtained from the Mikulski Archive for Space Telescopes MAST data archive at the Space Telescope Science Institute (STScI). Funding for the TESS mission is provided by the NASA Explorer Program. STScI is operated by the Association of Universities for Research in Astronomy, Inc., under NASA contract NAS 5–26555. We acknowledge the use of public TESS data from pipelines at the TESS Science Office and at the TESS Science Processing Operations Center.  Resources supporting this work were provided by the NASA High-End Computing (HEC) Program through the NASA Advanced Supercomputing (NAS) Division at Ames Research Center for the production of the SPOC data products.
This research has made use of the Exoplanet Follow-up Observation Program (ExoFOP; DOI: 10.26134/ExoFOP5) website, which is operated by the California Institute of Technology, under contract with the National Aeronautics and Space Administration under the Exoplanet Exploration Program.
M.G. and E.J. are FNRS-F.R.S. Research Directors. J.d.W. and MIT gratefully acknowledge financial support from the Heising-Simons Foundation, Dr. and Mrs. Colin Masson and Dr. Peter A. Gilman for Artemis, the first telescope of the SPECULOOS network situated in Tenerife, Spain. The ULiege's contribution to SPECULOOS has received funding from the European Research Council under the European Union's Seventh Framework Programme (FP/2007-2013) (grant Agreement n$^\circ$ 336480/SPECULOOS), from the Balzan Prize and Francqui Foundations, from the Belgian Scientific Research Foundation (F.R.S.-FNRS; grant n$^\circ$ T.0109.20), from the University of Liege, and from the ARC grant for Concerted Research Actions financed by the Wallonia-Brussels Federation. 
The research leading to these results has received funding from  the ARC grant for Concerted Research Actions, financed by the Wallonia-Brussels Federation. TRAPPIST is funded by the Belgian Fund for Scientific Research (Fond National de la Recherche Scientifique, FNRS) under the grant PDR T.0120.21. TRAPPIST-North is a project funded by the University of Liege (Belgium), in collaboration with Cadi Ayyad University of Marrakech (Morocco).
This work is based upon observations carried out at the Observatorio Astron\'omico Nacional on the Sierra de San Pedro M\'artir (OAN-SPM), Baja California, M\'exico.
SAINT-EX observations and team were supported by the Swiss National Science Foundation (PP00P2-163967, PP00P2-190080 and SPIRIT-216537),
the Centre for Space and Habitability (CSH) of the University of Bern,  the National Centre for Competence in Research PlanetS, supported by the Swiss National Science Foundation (SNSF). Y.G.M.C and A.K. are partially supported by UNAM PAPIIT-IG101321. B.-O.D. acknowledges support from the Swiss State Secretariat for Education, Research and Innovation (SERI) under contract number MB22.00046.
M.L. acknowledges support of the Swiss National Science Foundation under grant number PCEFP2\_194576.
Authors F.J.P and G.M. acknowledge financial support from the Severo Ochoa grant CEX2021-001131-S MICIU/AEI/10.13039/501100011033 and 
Ministerio de Ciencia e Innovación through the project PID2022-137241NB-C43 and the Ramón y Cajal grant RYC2022-037854-I.
A. P-T. acknowledges financial support from the Severo Ochoa grant CEX2021-001131-S funded by MICIU/AEI/ 10.13039/501100011033. 
F. M. acknowledges the financial support from the Agencia Estatal de Investigaci\'{o}n del Ministerio de Ciencia, Innovaci\'{o}n y Universidades (MCIU/AEI) through grant PID2023-152906NA-I00.}

\onecolumn
\section{Ground-Based Transit Observations}
\label{sec:ground-based-transit-obs}

\begin{table*}[ht!]
 \caption{Ground-based transit observation log.}
 \begin{center}
 {\renewcommand{\arraystretch}{1.1}
 \resizebox{0.84\textwidth}{!}{
 \begin{tabular}{l c c c c c}
 \toprule
Target & Telescope &  Date (UT) & Filter &  Exposure & Photometric   \\  
       &           &            &        & Time (s) & Aperture (") \\
 \hline 
TOI-1743.01 & RCO-0.4m & Jun 2 2020 & Sloan-$i'$ &180 & 3.7 \\
TOI-1743.01 & LCO-McD & Jul 7 2020 & Sloan-$i'$  &  82& 3.9  \\
TOI-1743.01 & TRAPPIST-N & Jul 19 2020  & $I+z'$  & 40& 4.8  \\
TOI-1743.01 & RCO-0.4m & Sept 4 2020 & Sloan-$i'$ & 180&5.1 \\
TOI-1743.01 & TRAPPIST-N & Sept 21 2020  & $I+z'$  &40 & 7.2 \\
TOI-1743.01 & KeplerCam & Apr 10 2021 & Sloan-$i'$ & 60&4.7 \\
TOI-1743.01 & MuSCAT3 & May 27 2021 & Sloan-$g'$,-$r'$,-$i'$, $z_s$  & 180,55,28,27 & 3.0  \\
TOI-1743.01 & TUG-T100 & Oct 29 2022 & Sloan-$i'$ & 45 &5.6 \\
TOI-1743.01 & SAINT-EX & May 5 2024 & Sloan-$z'$ & 17 &4.9 \\
TOI-1743.01 & TUG-T100 & Aug 3 2024 & Sloan-$i'$ & 50& 4.8 \\
\hline
TOI-5799.01 & TRAPPIST-N & Sept 28 2022  & Sloan-$z'$  & 15 &4.8  \\
TOI-5799.01 & LCO-Teid & Jun 14 2023 & Sloan-$i'$  &22 & 3.9  \\
TOI-5799.01 & MuSCAT2 & Jul 8 2023 & Sloan-$g'$,-$r'$,-$i'$, $z_s$ &10,10,5,5 &10.8 \\
TOI-5799.01 & TUG-T100 & Aug 23 2023 & Sloan-$g'$  & 120 &5.5  \\
TOI-5799.01 & TUG-T100 & Aug 11 2024 & Sloan-$z'$  & 50 &6.1  \\
TOI-5799.01 & TUG-T100 & Sept 26 2024 & Sloan-$g'$  & 30 &2.8  \\
TOI-5799.02 & TUG-T100 & Oct 23 2024 & Sloan-$g'$  & 75& 5.5  \\
TOI-5799.02 & AUKR-T80 & Oct 23 2024 & Sloan-$i'$  &  50&5.4  \\
TOI-5799.01 & Artemis & Oct 25 2024  & Sloan-$r'$  & 10 &3.5  \\
\hline
TOI-6223.01 &  TUG-T100 & Aug 23 2023 & Sloan-$i'$  & 120 &10.0 \\
TOI-6223.01 &  Artemis & Jul 27 2024 & Sloan-$z'$  & 16 &2.5 \\
TOI-6223.01 &  TUG-T100 & Jul 31 2024 & Sloan-$g'$  & 100 &4.5 \\
TOI-6223.01 &  MuSCAT2 & Aug 23 2024 & Sloan-$g'$,-$r'$,-$i'$, $z_s$  & 90,90,15,80 &10.9 \\
TOI-6223.01 &  MuSCAT2 & Sept 23 2024 & Sloan-$g'$,-$r'$,-$i'$, $z_s$  & 90,90,15,80& 10.9 \\
TOI-6223.01 &  Artemis & Oct 20 2024 & Sloan-$r'$  & 45 &1.8 \\
TOI-6223.01 &  TUG-T100 & Oct 20 2024 & Sloan-$g'$  & 100 &6.2 \\
 \hline
 \end{tabular}}}
 \end{center}
 \label{obs_table}
\end{table*}

\begin{figure}[!]
	\centering
	\includegraphics[width=0.49\columnwidth]{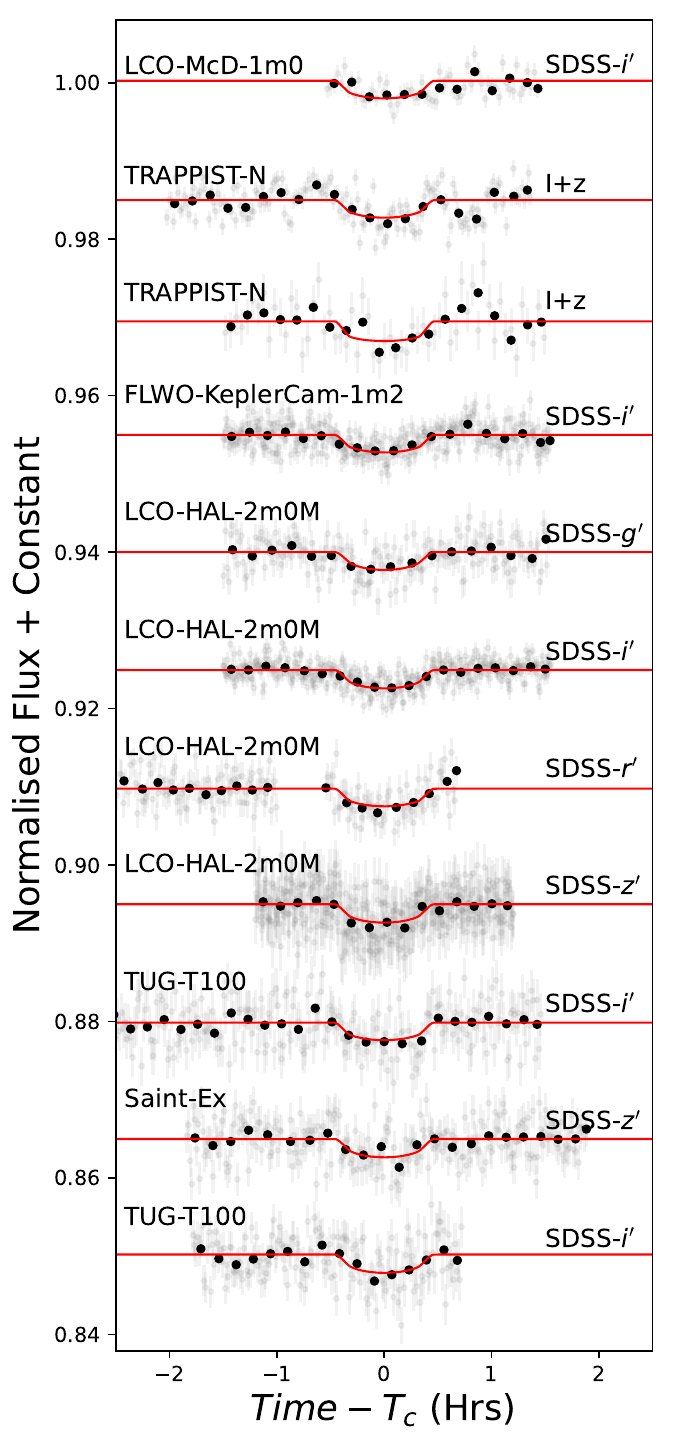}
    \includegraphics[width=0.49\columnwidth]{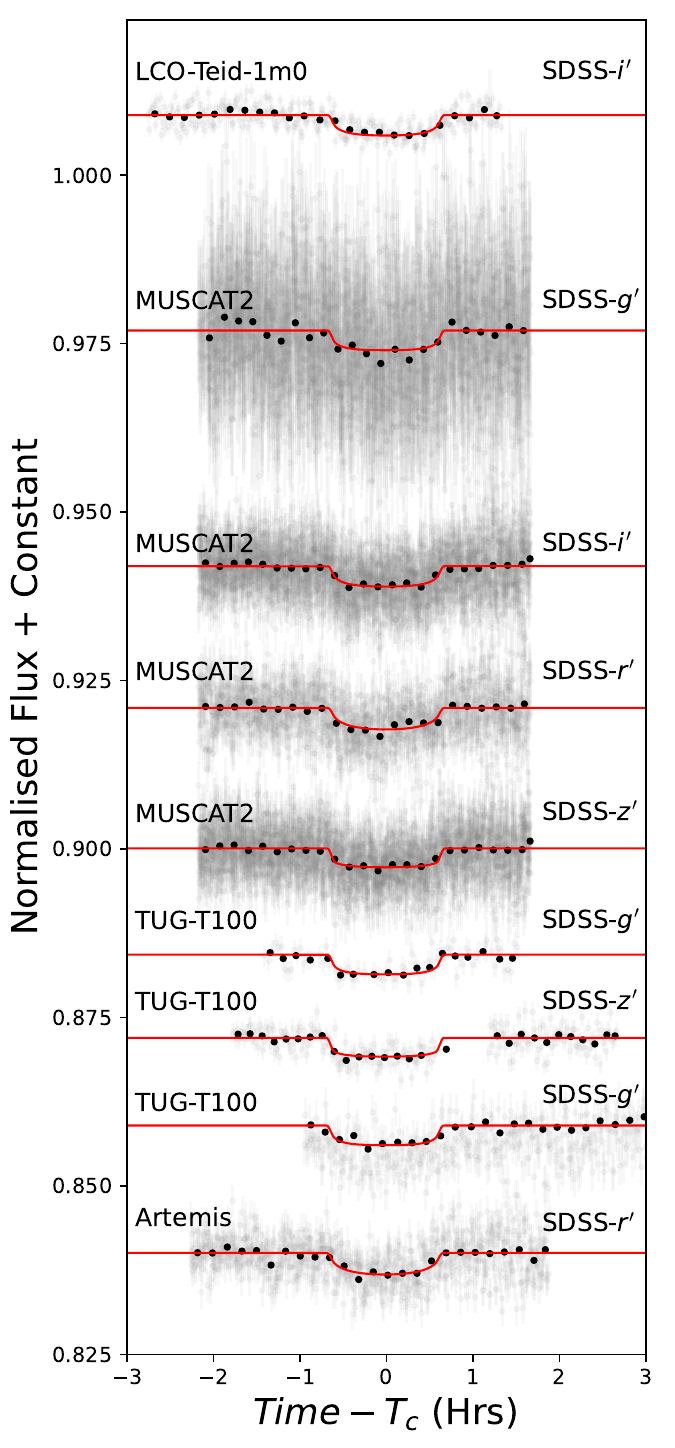}
	\caption{Ground-based transit observations of TOI-1743\,b (left) and TOI-5799\,b (right) shown with gray dots with error bars. Black dots are the 10 minute bins while red continious lines are the global model.}
	\label{fig:ground_based_obs_1743_5799b}
\end{figure}

\twocolumn
\begin{figure}
	\centering
    \includegraphics[width=\columnwidth]{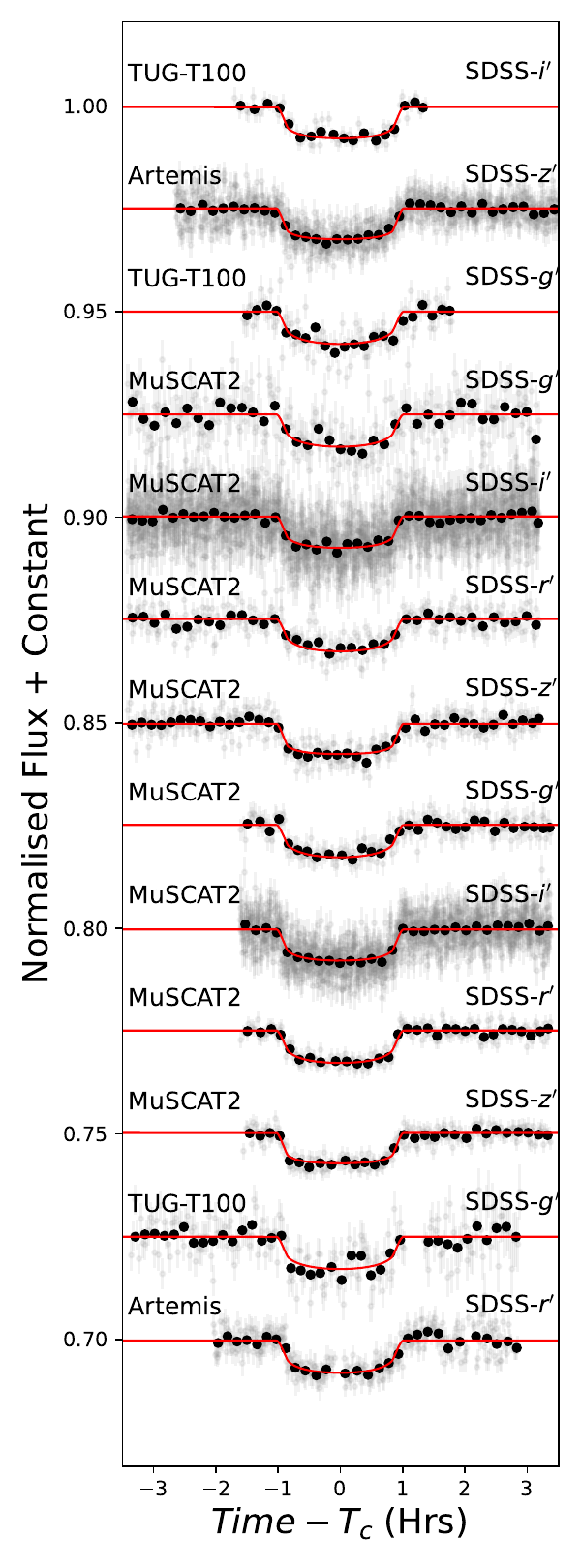}
	\caption{Same as Fig \ref{fig:ground_based_obs_1743_5799b} but for  TOI-6223\,b}.
	\label{fig:ground_based_obs_5938_6223}
\end{figure}

\begin{figure}
	\centering
	\includegraphics[width=1.0\columnwidth]{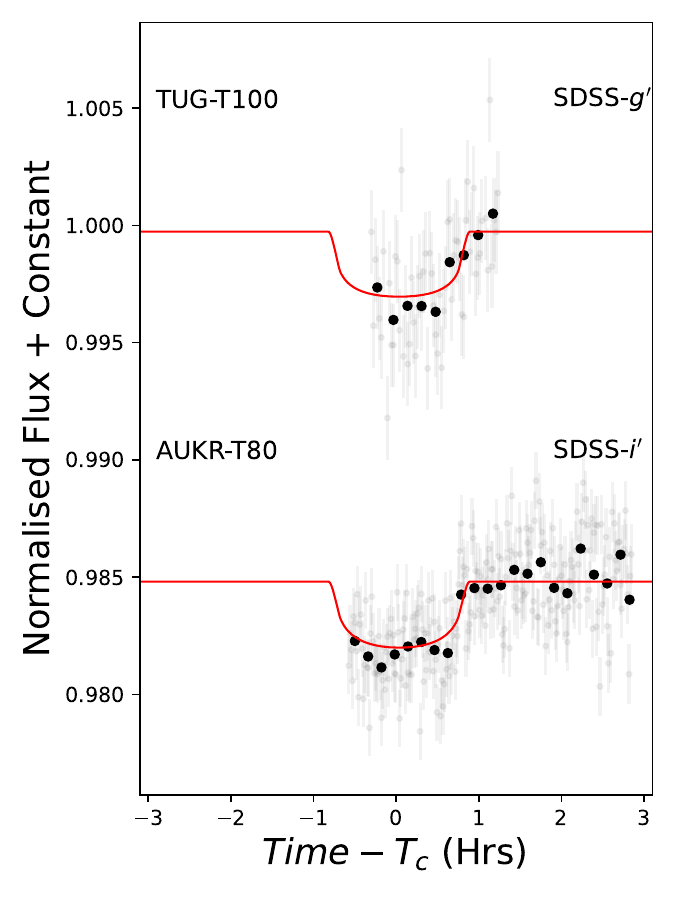}
	\caption{Same as Fig \ref{fig:ground_based_obs_1743_5799b} but for TOI-5799\,c.}
	\label{fig:ground_based_obs5799c}
\end{figure}

\twocolumn
\section{SED Models}\label{sec:SED Models}

\begin{figure}[ht!]
    \centering
    \includegraphics[width=1\columnwidth, keepaspectratio]{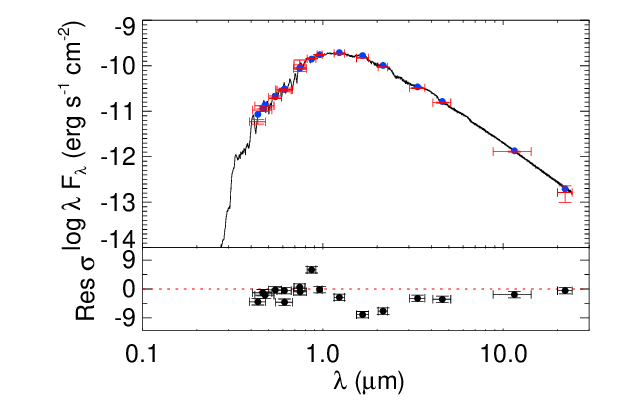}
    \includegraphics[width=1\columnwidth, keepaspectratio]{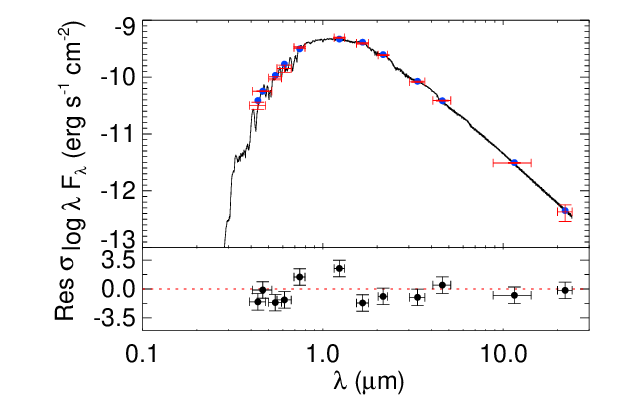}
    \includegraphics[width=1\columnwidth, keepaspectratio]{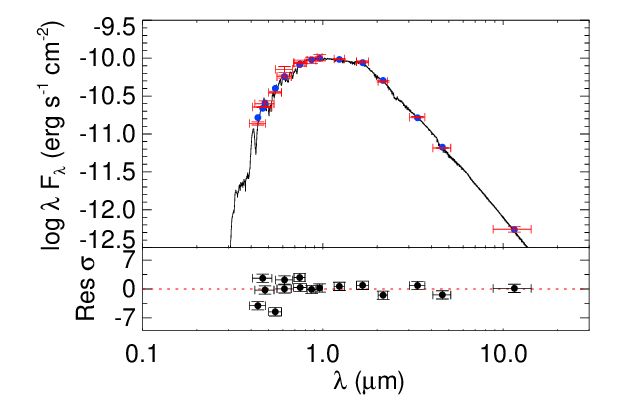}
    \caption{
        SED models (black lines) of host stars (from top to bottom: TOI-1743, TOI-5799, and TOI-6223). Red dots with error bars represent the broadband fluxes, while their corresponding model values are shown with blue dots. Horizontal red bars shows the bandwidth of filters. Residuals are shown in lower panel for each plot in $\sigma$ units. \\ \\ \\ \\
}
    \label{fig:SED}
\end{figure}

\section{Simulated JWST Transmission Spectra}\label{sec:Simulated JWST Transmission Spectra}

\begin{figure*}
    \centering
    \includegraphics[width=1.1\columnwidth]{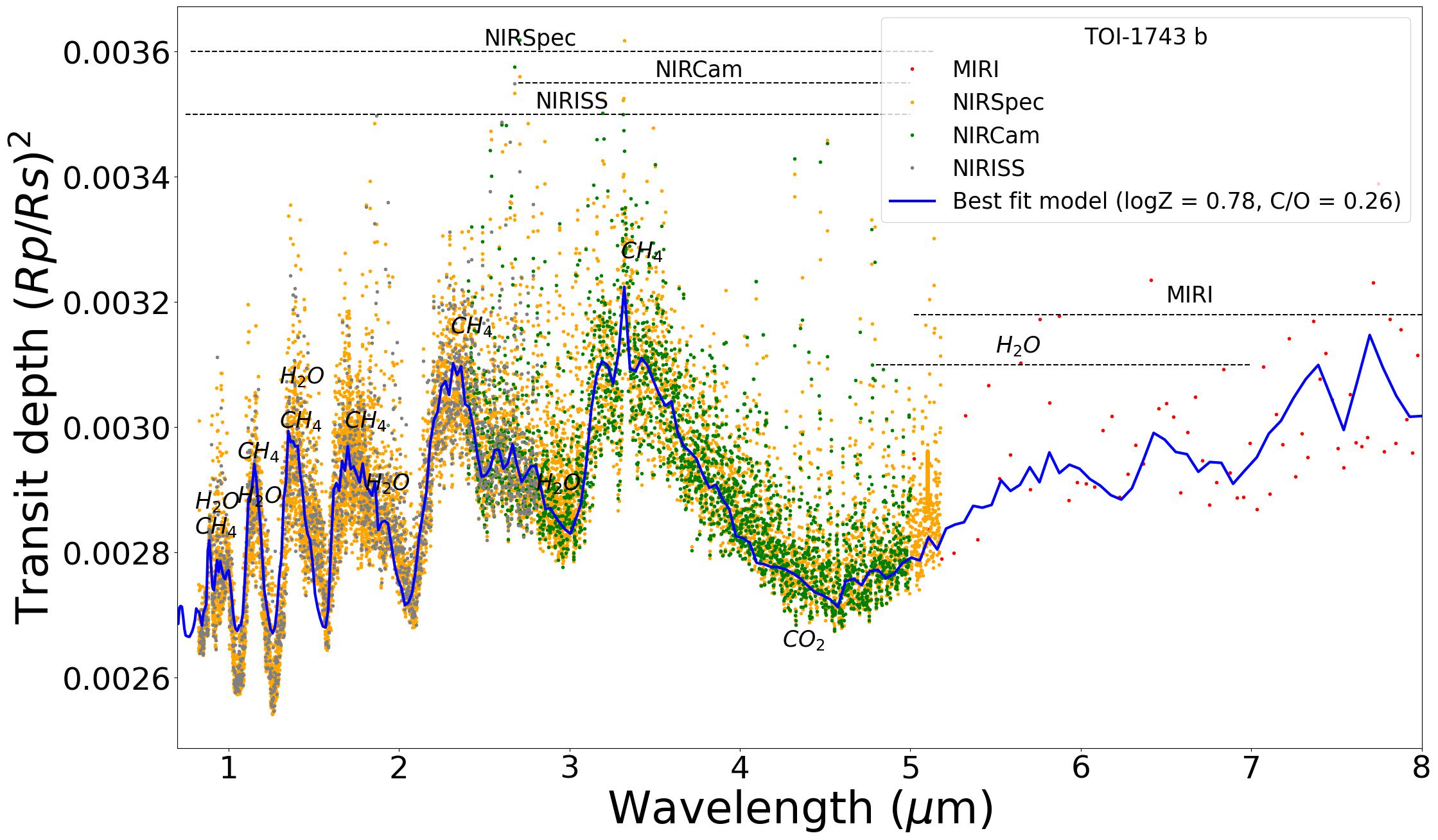}
    \includegraphics[width=1.1\columnwidth]{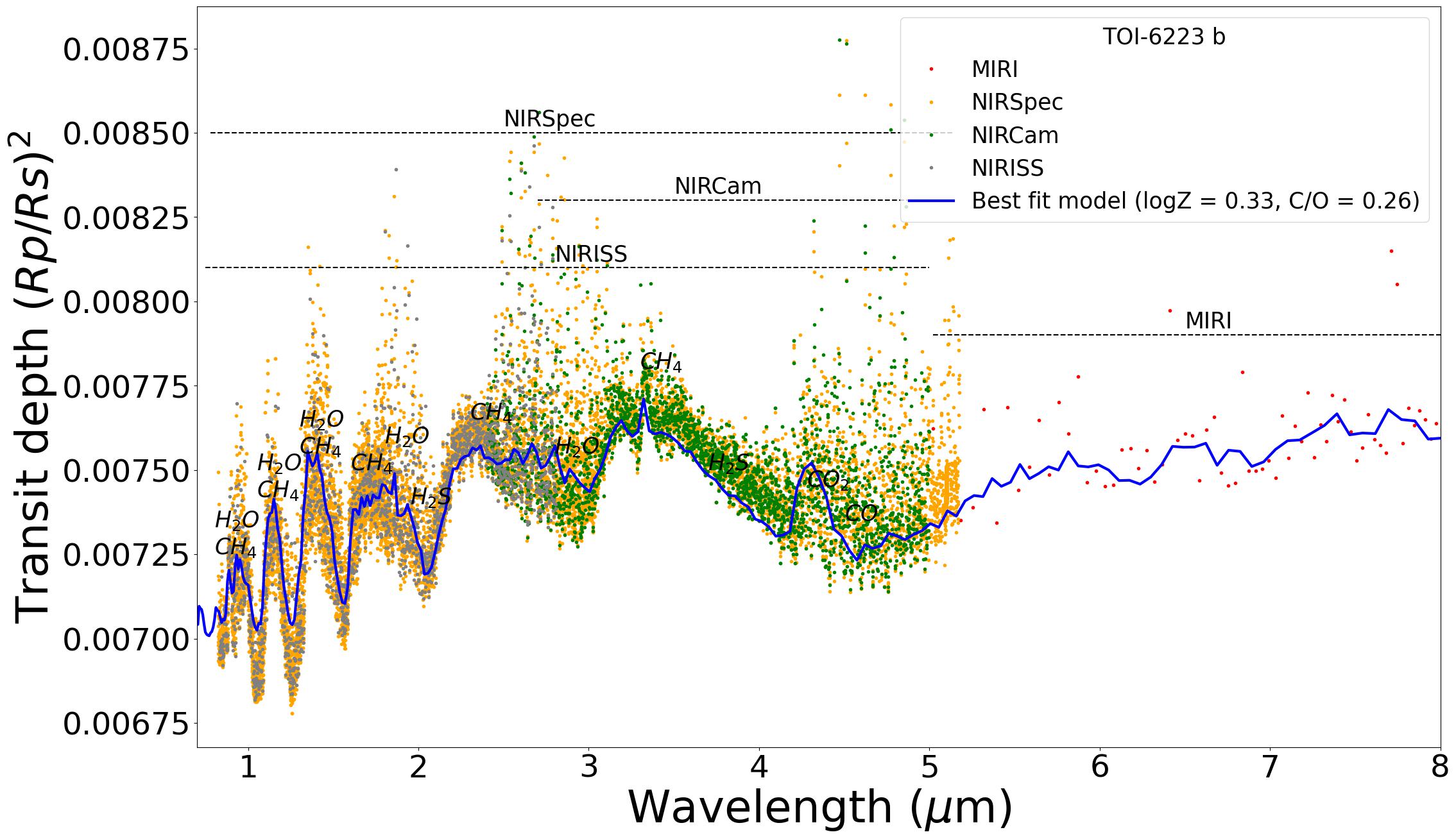}
    \includegraphics[width=1.1\columnwidth]{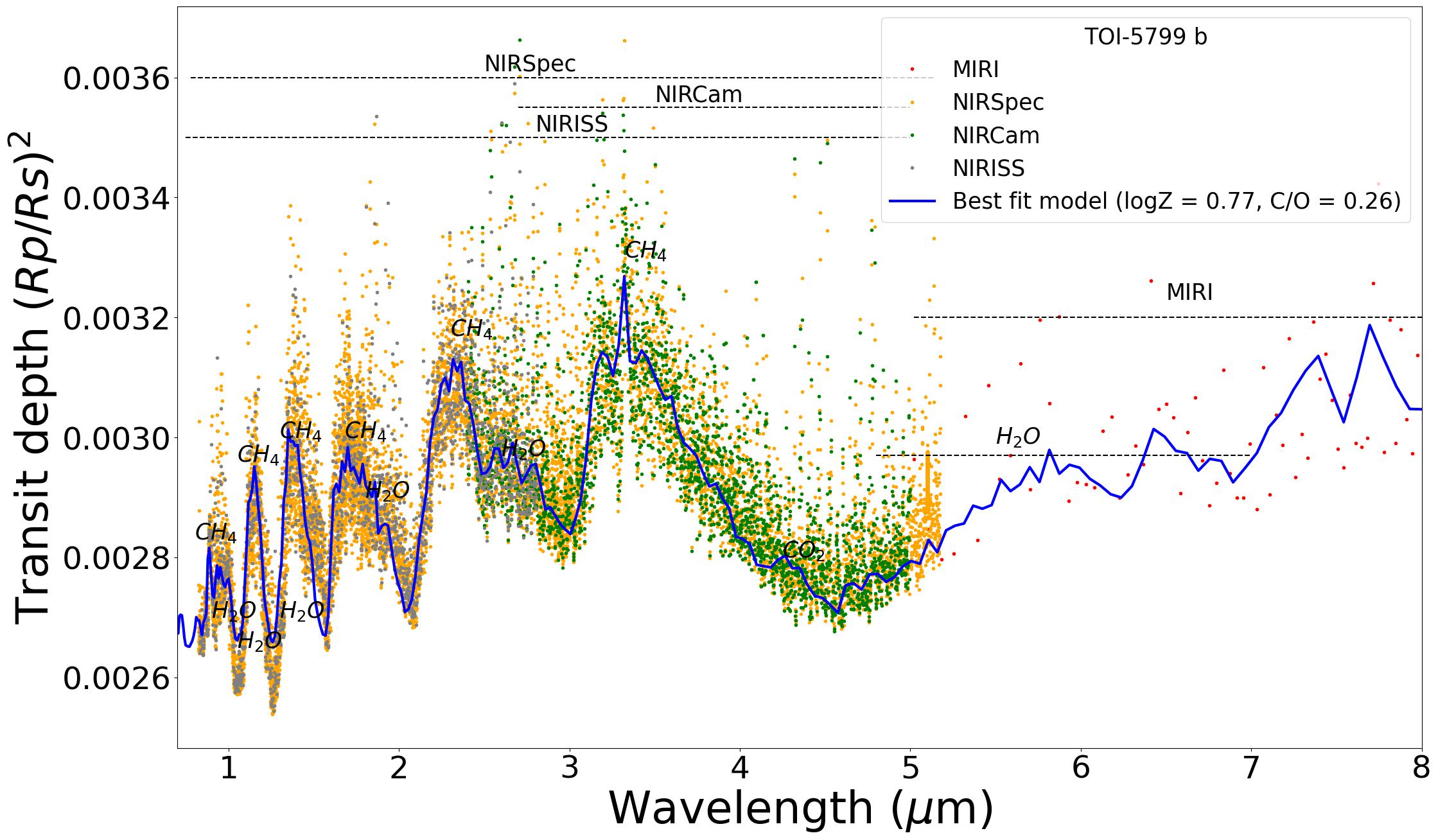}
    \includegraphics[width=1.1\columnwidth]{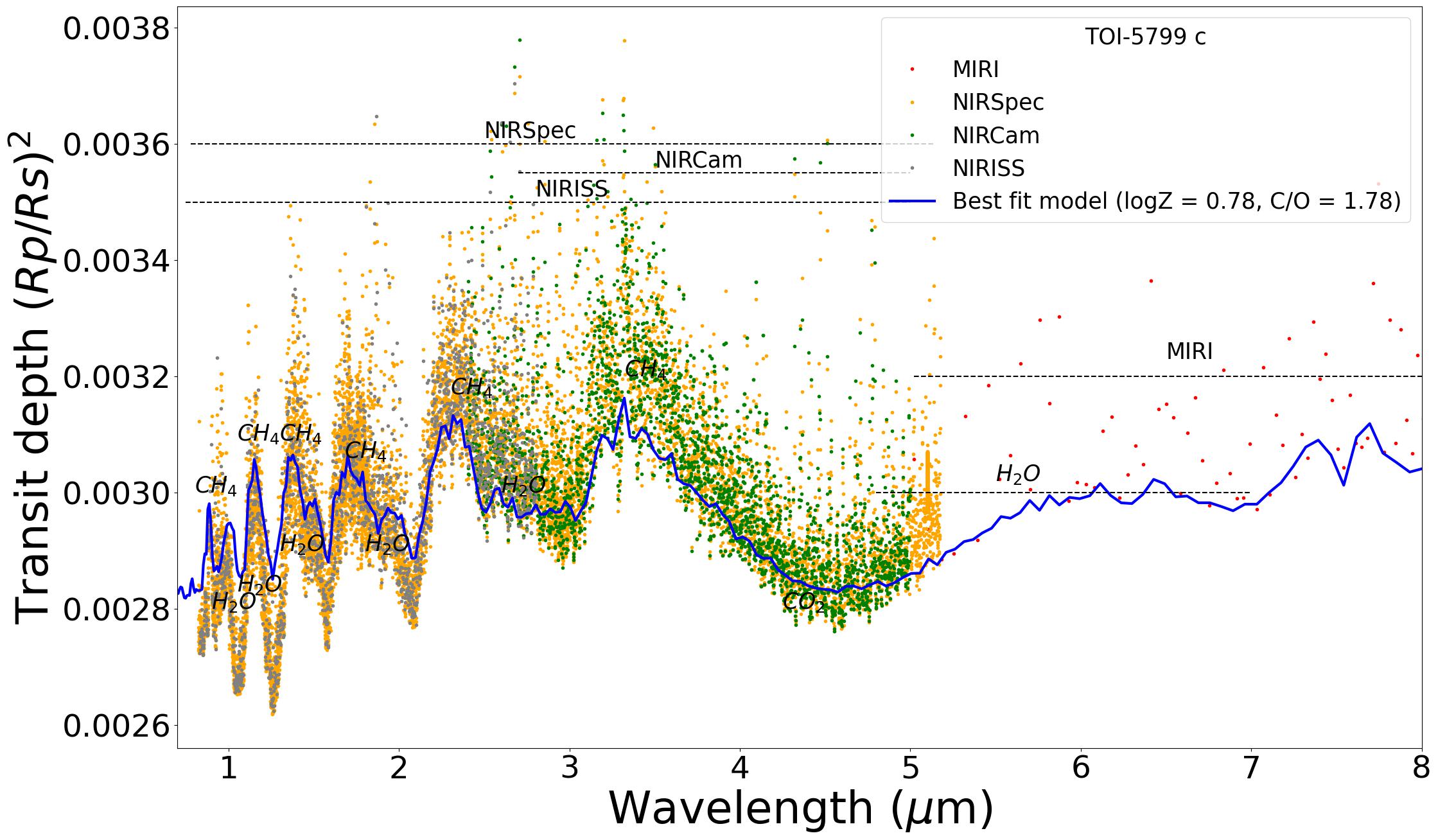}
    \caption{Synthetic transmission spectra from \emph{JWST} for TOI-1743\,b, TOI-5799\,b, TOI-5799\,c and TOI-6223\,b generated with the help of the \texttt{PLATON} and \texttt{PandExo} tools. Blue line shows the best-fit model on the synthetic data simulated using JWST's MIRI, NIRSpec, NIRCam, and NIRISS instruments.}
    \label{fig:STS}
\end{figure*}

\begin{table}[ht!]
\caption{Exposure times and S/N obtained for each synthetic spectrum of JWST.}
    \centering
    {\renewcommand{\arraystretch}{1.}
    \begin{tabular}{c c c c}
         \hline
         Instrument & Mode$^a$  & Exposure & S/N \\
                    &           & Time     &     \\
         \hline
         \hline
         \multicolumn{4}{c}{\bf TOI-1743\,b} \\
         MIRI & Slit Spectroscopy & 930 s & 400 \\
         NIRSpec & G140H/F070LP$^b$ & 1115 s & 1195 \\
                 & G140H/F100LP & 1115 s & 1165 \\
                 & G235H/F170LP & 1135 s & 1225 \\
                 & G395H/F290LP & 1150 s & 1050 \\
         NIRCam & F322W2 & 1145 s & 1115 \\
                & F444W  & 1135 s & 850 \\
         NIRISS & SUBSTRIP256$^c$ & 1155 s & 1680 \\
         \multicolumn{4}{c}{\bf TOI-5799\,b} \\
         MIRI & Slit Spectroscopy & 1080 s & 805 \\
         NIRSpec & G140H/F070LP & 1630 s & 2265 \\
                 & G140H/F100LP & 1630 s & 2300 \\
                 & G235H/F170LP & 1625 s & 2270 \\
                 & G395H/F290LP & 1630 s & 1930 \\
         NIRCam & F322W2 & 1650 s & 2160 \\
                & F444W  & 1660 s & 1620 \\
         NIRISS & SUBSTRIP256 & 1650 s & 3115 \\
         \multicolumn{4}{c}{\bf TOI-5799\,c} \\
         MIRI & Slit Spectroscopy & 1390 s & 910 \\
         NIRSpec & G140H/F070LP & 2085 s & 2560 \\
                 & G140H/F100LP & 2090 s & 2600 \\
                 & G235H/F170LP & 2085 s & 2570 \\
                 & G395H/F290LP & 2095 s & 2185 \\
         NIRCam & F322W2 & 2095 s & 2435 \\
                & F444W  & 2085 s & 1820 \\
         NIRISS & SUBSTRIP256 & 2080 s & 3495 \\
         \multicolumn{4}{c}{\bf TOI-6223\,b} \\
         MIRI & Slit Spectroscopy & 2260 s & 650 \\
         NIRSpec & G140H/F070LP & 2465 s & 1340 \\
                 & G140H/F100LP & 2440 s & 1370 \\
                 & G235H/F170LP & 2435 s & 1245 \\
                 & G395H/F290LP & 2475 s & 1010 \\
         NIRCam & F322W2 & 2560 s & 1165 \\
                & F444W  & 2460 s & 795 \\
         NIRISS & SUBSTRIP256 & 2522 s & 1895 \\
         \hline
    \end{tabular}}
    $^a$Information of the different modes, resolving power and wavelength can be found at: \url{https://jwst-docs.stsci.edu/}\\
    $^b$We chose various disperser-filter combinations that can provide a high resolution power of $~$2700. G = Grism, F = Filter.\\
    $^c$Size of the subarray for single-object slitless spectroscopy. \\

    \label{tab:ExpSNR}
\end{table}

\clearpage
\twocolumn
\section{High-resolution Observations}\label{sec:High-resolution Observations}

\begin{figure}[h]
	\centering
    \includegraphics[scale=0.4]{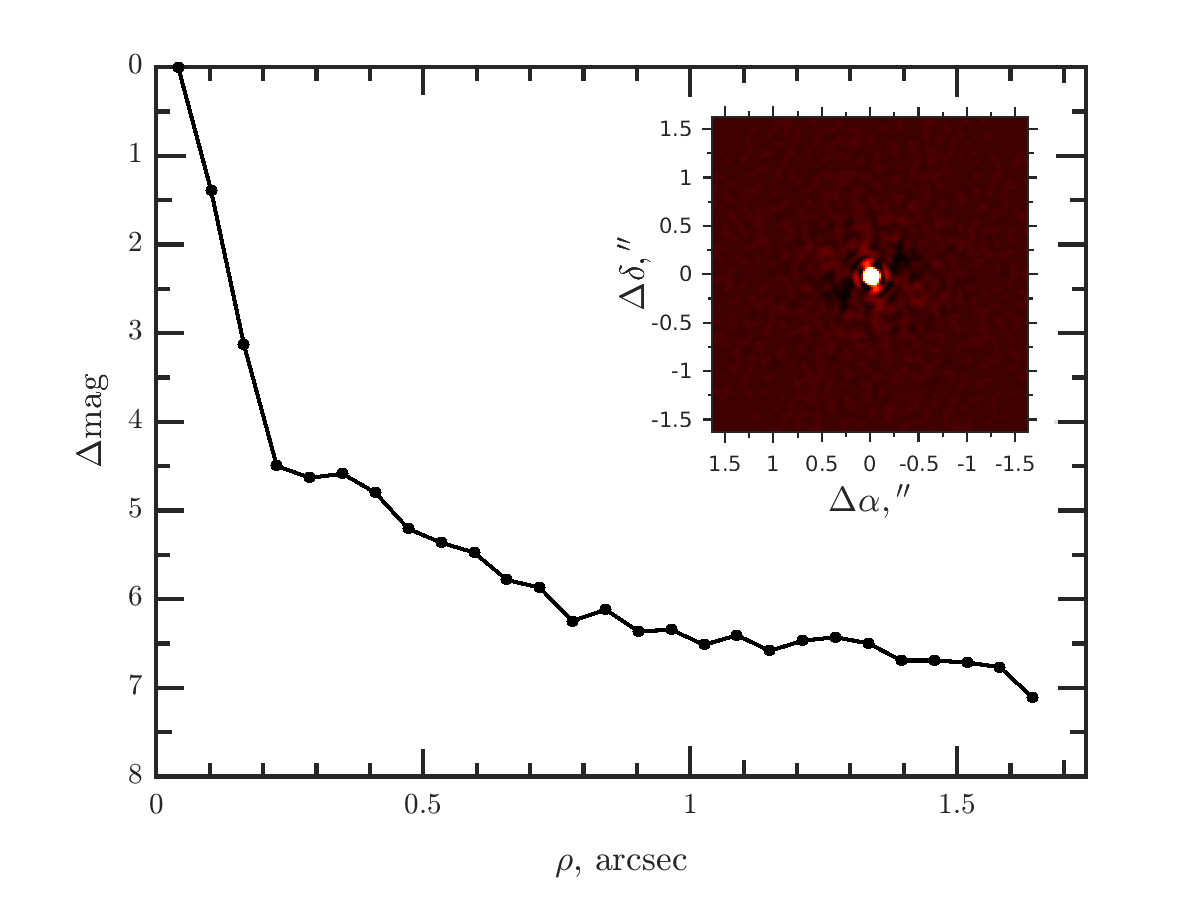}
	\caption{High-resolution imaging TOI-6223 from SAI and contrast curve as a function of angular separation. No stellar companions were found within detection limits for both targets.}
	\label{fig:high_res_SAI}
\end{figure}

\begin{figure}[h]
	\centering
    \includegraphics[scale=0.5]{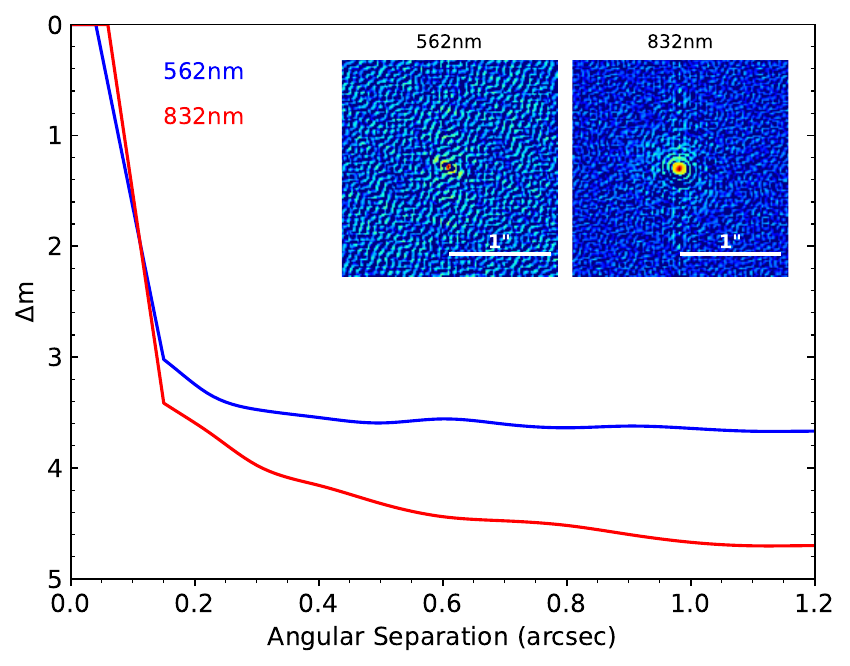}
    \includegraphics[scale=0.5]{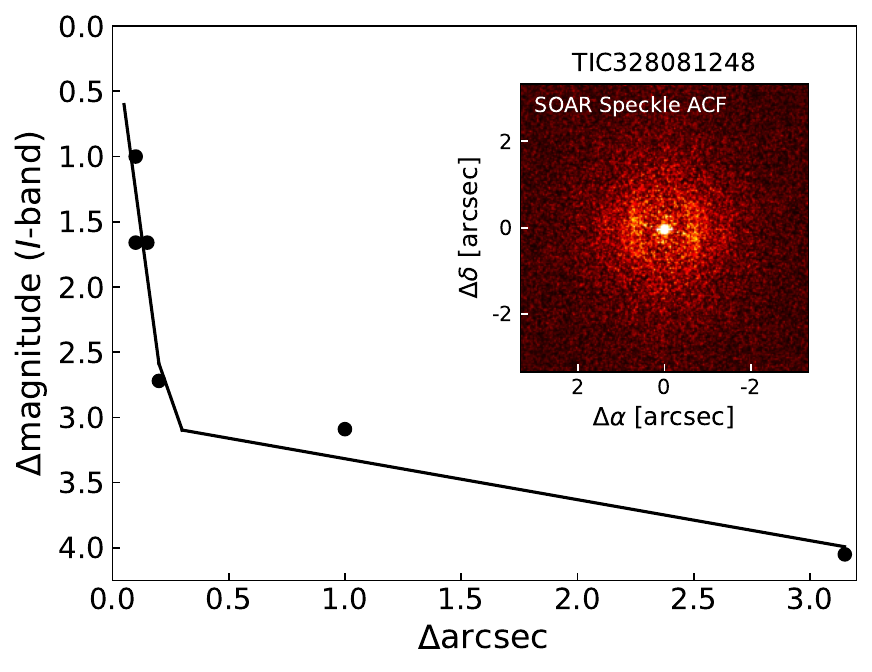}
	\caption{High-resolution images and contrast curves of TOI-1743 from WIYN (top panel) and TOI-5799 from SOAR (bottom panel). Contrast curve of TOI-1743 in blue filter centered at 562 nm shown in blue line while the red filter centered at 832 nm shown with red line. Contrast curve of TOI-5799 in I filter is shown with black line. No additional source was found in both images.}
	\label{fig:high_res_WIYN}
\end{figure}

\begin{figure}[h]
	\centering
    \includegraphics[scale=0.09]{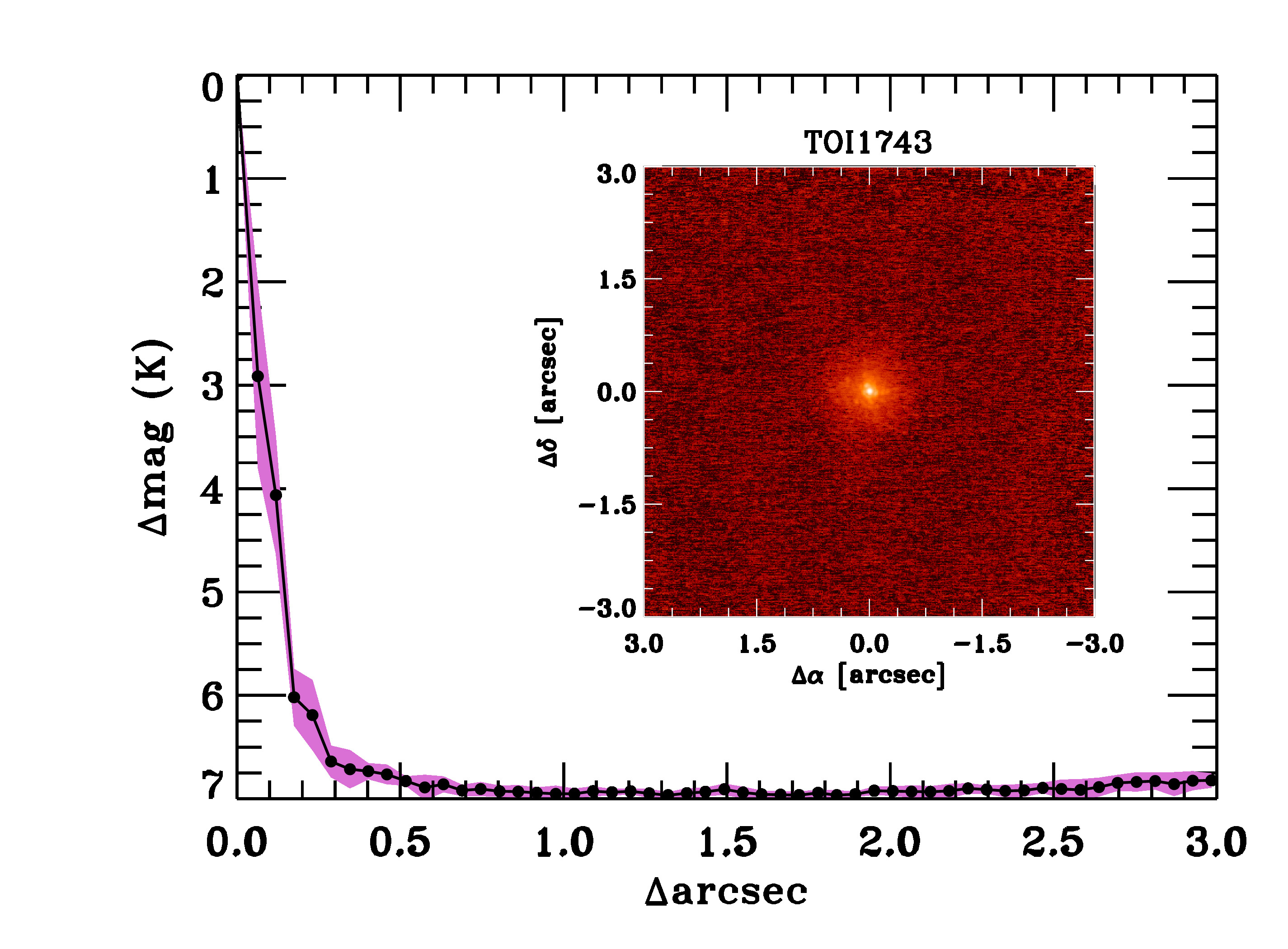}
    \includegraphics[scale=0.09]{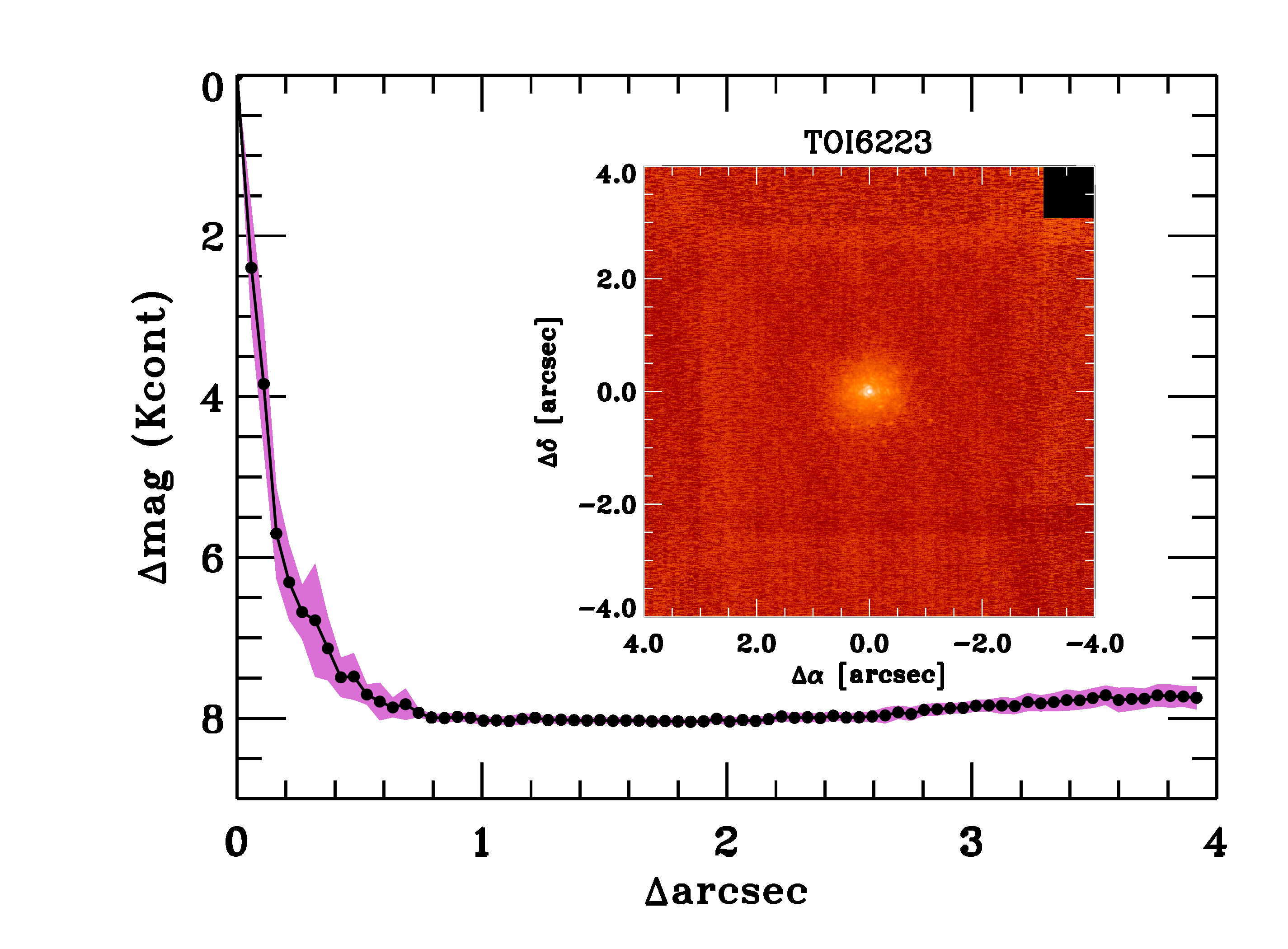}
	\caption{High-resolution imaging of TOI-1743 in K filter (top panel) and TOI-6223 in $K_c$ filter (bottom panel) with contrast curve from Keck. No additional source was found near both stars.}
	\label{fig:high_res_Keck}
\end{figure}

\end{document}